\documentclass[sigconf]{acmart}
 
\AtBeginDocument{%
  }

\copyrightyear{2025}
\acmYear{2025}
\setcopyright{cc}
\setcctype{by}
\acmConference[UIST '25]{The 38th Annual ACM Symposium on User Interface Software and Technology}{September 28-October 1, 2025}{Busan, Republic of Korea}
\acmBooktitle{The 38th Annual ACM Symposium on User Interface Software and Technology (UIST '25), September 28-October 1, 2025, Busan, Republic of Korea}\acmDOI{10.1145/3746059.3747612}
\acmISBN{979-8-4007-2037-6/2025/09}
\acmISBN{978-1-4503-XXXX-X/2018/06}

\newcommand{\ipstart}[1]{\vspace{1mm}\noindent{\textbf{\textit{#1.}}}}

\newcommand{\camready}[1]{\textcolor{black}{#1}}

\newcommand{\sysname}{Vid2Coach}
\usepackage{multirow}
\usepackage{mdframed}
\usepackage{makecell}
\usepackage{array} 

\begin{document}


\title{Vid2Coach: Transforming How-To Videos into Task Assistants}

\author{Mina Huh}
\affiliation{
  \institution{The University of Texas at Austin}
  \country{Austin, Texas, USA}}
\email{minahuh@cs.utexas.edu}

\author{Zihui Xue}
\affiliation{
  \institution{The University of Texas at Austin}
  \country{Austin, Texas, USA}}
\email{sherryxue@utexas.edu}

\author{Ujjaini Das}
\affiliation{
  \institution{The University of Texas at Austin}
  \country{Austin, Texas, USA}}
\email{ujjaini@utexas.edu}

\author{Kumar Ashutosh}
\affiliation{
  \institution{The University of Texas at Austin}
  \country{Austin, Texas, USA}}
\email{kumar.ashutosh@utexas.edu}

\author{Kristen Grauman}
\affiliation{
  \institution{The University of Texas at Austin}
  \country{Austin, Texas, USA}}
\email{grauman@cs.utexas.edu}

\author{Amy Pavel}
\affiliation{
  \institution{University of California, Berkeley}
  \country{Berkeley, California, USA}}
\email{amypavel@eecs.berkeley.edu}

\begin{abstract}
People use videos to learn new recipes, exercises, and crafts. Such
videos remain difficult for blind and low vision (BLV) people to
follow as they rely on visual comparison. Our observations of visual rehabilitation therapists (VRTs) guiding BLV people to follow how-to videos revealed that VRTs provide both proactive and responsive support including detailed descriptions, non-visual workarounds, and progress feedback. We propose Vid2Coach, a system that transforms how-to videos into wearable camera-based assistants that provide accessible instructions and mixed-initiative feedback. From the video, Vid2Coach generates accessible instructions by augmenting narrated instructions with demonstration details and completion criteria for each step. It then uses retrieval-augmented-generation to extract relevant non-visual workarounds from BLV-specific resources. 
Vid2Coach then monitors user progress with a camera embedded in commercial smart glasses to provide context-aware instructions, proactive feedback, and answers to user questions.
BLV participants (N=8) using Vid2Coach completed cooking tasks with 58.5\% fewer errors than when using their typical workflow and wanted to use Vid2Coach in their daily lives. Vid2Coach demonstrates an opportunity for AI visual assistance that strengthens rather than replaces non-visual expertise.
\end{abstract}


\begin{teaserfigure}
  \includegraphics[width=\textwidth]{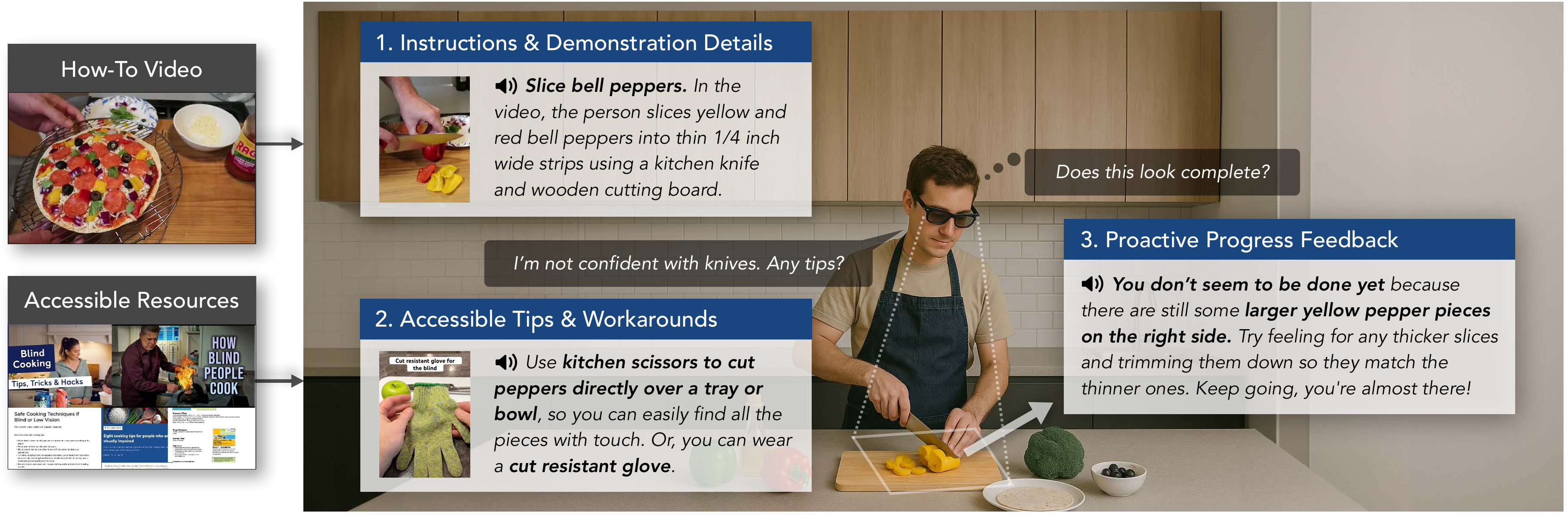}
  \caption{Vid2Coach is a system that transforms how-to videos into a wearable camera-based task assistant that provides accessible instructions and mixed-initiative feedback. Given a how-to video, Vid2Coach extracts high-level steps and demonstration details, then for each step we use retrieval-augmented generation to supplement each step with BLV-specific guidelines. Vid2Coach then monitors user progress with a camera in smart glasses to provide proactive feedback.}
  \label{fig:teaser}
\end{teaserfigure}

\maketitle
\section{Introduction}
People watch how-to videos to learn new skills and follow everyday procedural tasks such as cooking~\cite{chang2021rubyslippers}, exercising~\cite{rector2013eyes}, makeup~\cite{li2022feels, truong2021automatic}, and crafting~\cite{chi2013democut}. 
Videos support efficient skill learning~\cite{ostrow2014testing, prihar2023investigating} as their rich visual demonstrations share tacit knowledge, low-level details,
and allow learners to visually compare their progress.
The meteoric rise of video platforms such as YouTube~\cite{youtube} and TikTok~\cite{tiktok} has led to the proliferation of free and high-quality how-tos, and to many creators sharing knowledge \textit{only} via video. However, blind and low vision (BLV) learners who want to acquire new skills through videos~\cite{li2022feels, krolak2017accessibility, seo2018understanding} find most videos prohibitively difficult to follow~\cite{li2024recipe, liu2021makes, liu2022crossa11y} 
as videos typically omit key visual details in their narrations and include instructions that assume sight (\textit{e.g.}, \textit{``wait until your butter turns this color''}). 
Thus, BLV learners often learn new skills via 1:1 support, group classes, or text instructions. 1:1 support is valuable but rare, as it requires the instructor to know both how to perform the task and how to perform it non-visually. Specialized group classes teach non-visual approaches for common skills (\textit{e.g.}, how to use a stove) but they do not provide on-demand procedural learning to meet individual needs (\textit{e.g.}, repair a specific washer). Text instructions often lack non-visual procedures and provide less detail than videos.
Making videos meaningfully useful for BLV learners would address a longstanding learning gap. 

We envision a system that transforms how-to videos into wearable camera-based assistants for BLV learners. Our key insight is that the rich audio-visual content that benefits sighted learners can provide an AI assistant with detailed task knowledge to create accessible instructions and assess user progress. Further, new commercially available smart glasses offer hands-free capture of hand-object interactions required for real-time feedback, and many BLV individuals already use such technology in daily life~\cite{rayban_blind}. We approach this goal with cooking as an initial target domain due to its universality, multi-step complexity, diverse tools, and reliance on multimodal signals (vision, tactile, auditory) that create challenges and opportunities for accessible instruction.

To inform the design of our system, we first conducted an observational study with three pairs of vision rehabilitation therapists (VRTs) and BLV participants. VRTs remotely guided BLV participants to follow a cooking how-to video as BLV participants performed the task while streaming an egocentric video from their smart glasses. 
VRTs provided verbal instructions grounded in both the narration and visual cues from how-to videos -- describing tools, ingredients, and actions in detail. They also offered safety tips and accessible workarounds (\textit{e.g.,} using a talking scale) and encouraged BLV people to use sensory cues like touch, smell, and sound to assess the progress (\textit{e.g., ``Feel for any eggshells left in the bowl.''}). 
They provided both proactive and responsive visual descriptions tailored to each individual's vision level, cooking experience, and kitchen setup.
BLV participants valued receiving 1:1 support to learn new skills, but it was difficult for VRTs to monitor both the video and the user's progress and such VRT support is not scalable to millions of BLV learners. 



We present Vid2Coach (Figure~\ref{fig:teaser}),  a system that transforms how-to videos into wearable camera-based assistants that provide accessible instructions and mixed-initiative feedback. From a how-to video, Vid2Coach creates \textit{accessible instructions} by first extracting high-level steps then adding detailed demonstration descriptions and completion criteria for each step via multimodal understanding (Figure~\ref{fig:teaser}.1). To supplement instructions with tips and workarounds tailored to users' level of vision, cooking skills, and kitchen setup, our pipeline uses retrieval-augmented generation (RAG) to generate suggestions grounded in an accessibility resources dataset (Figure~\ref{fig:teaser}.2). 
As a user performs the task, Vid2Coach captures the egocentric video stream (Figure~\ref{fig:teaser}.3) to provide real-time \textit{mixed-initiative feedback}. Using a vision-language model (Gemini~\cite{team2023gemini}), it determines whether the user is following the current step or engaging in spontaneous, user-initiated actions (\textit{e.g.,} washing hands, clearing up the cutting board). When a step appears complete based on the completion criteria extracted from the how-to video, Vid2Coach notifies the user -- explaining visually grounded reasoning (\textit{e.g., ``You seem to be done because the bacon has turned evenly golden brown and crispy.''}) and asks whether they would like to move on. For steps where visual cues alone are insufficient to determine completion (\textit{e.g.,} mixing until smooth), it prompts users to use non-visual cues (touch, smell, or sound). At any point, users can interact via free-form voice input to repeat instructions, navigate between steps, or ask questions to receive responses grounded in both the how-to video and task-specific accessibility resources.

To evaluate Vid2Coach, we conducted a within-subjects study with 8 BLV people in their own home kitchens, where they completed end-to-end recipes. We compared Vid2Coach to users' current approach for accessing how-to videos (\textit{e.g.,} original video and transcript) and getting feedback (\textit{e.g.,} chatGPT~\cite{achiam2023gpt}, BeMyEyes~\cite{BeMyEyes}, AIRA~\cite{Aira}). Participants using Vid2Coach completed cooking tasks with 58.5\% fewer errors and reported lower cognitive load. 
To demonstrate Vid2Coach's applicability beyond cooking, we conducted an exploratory extension study that involved assembly and decoration tasks, revealing key considerations to support a wider range of hands-on activities. 

The contributions of our paper are as follows:
\begin{itemize}
    \item Observational study that reveals how VRTs provide remote guidance to BLV individuals following a procedural task
    \item \textit{Vid2Coach}, a system that transforms how-to videos into an accessible task assistant
    \item Automatic pipeline for generating instructions from how-to videos and providing real-time feedback on user progress, validated with evaluation
    \item User study showcasing the advantages and limitations of using Vid2Coach
\end{itemize}


\section{Related Work}
\subsection{Video Accessibility}
Learning new tasks from videos is challenging partially due to the inaccessibility of the videos themselves. 
While BLV people are interested in learning new skills from videos such as cooking~\cite{li2021non, li2024recipe}, exercises~\cite{rector2013eyes}, and makeup~\cite{li2022feels}, the visual content is not fully described in narration~\cite{liu2021makes, liu2022crossa11y, peng2021say}. 
To make videos accessible, volunteers~\cite{YouDescribe} or professional audio describers~\cite{pavel2020rescribe} add ~\textit{audio descriptions}, which verbally describe actions, characters, or settings.
Because creating audio descriptions is time-consuming, prior research proposed authoring tools to facilitate the process by identifying undescribed visual elements~\cite{peng2021say, liu2022crossa11y}, and optimizing the placement of descriptions~\cite{pavel2020rescribe}.
Beyond manual creation of audio descriptions, researchers have also explored automatic generation of audio descriptions to support video accessibility at scale~\cite{wang2021toward, li2025videoa11y}. These systems can automatically describe scene changes~\cite{huh2023avscript, peng2021slidecho}, provide hierarchical descriptions for interactive access~\cite{van2024making, chang2024worldscribe, ning2024spica}, and synthesize speech of audio description script~\cite{MS_AIAD}. 
As BLV people have varied preferences across viewing scenarios~\cite{jiang2024s}, prior work proposed systems that enable users to customize the length and speech styles of audio descriptions~\cite{natalie2024audio} and prioritize visual elements that are more commonly emphasized by other viewers~\cite{xu2025danmua11y, huh2022cocomix}. 

While these approaches enhance access to visual content in videos, they provide a high-level understanding of scenes in entertainment or social contexts (\textit{e.g.,} watching movies) rather than delivering detailed, task-oriented descriptions for following procedural tasks. As a result, they often omit information necessary for following instructions (\textit{e.g.,} how actions are performed or which tools are used) because such details are considered to interrupt the story flow or are compressed due to time constraints. 
In this work, we investigate how to improve the accessibility of how-to videos, enabling BLV individuals to more effectively learn and follow tasks through video content. We take an alternative approach by extracting task-relevant frames and generating fine-grained descriptions to support the understanding and evaluation of task progress.


\subsection{Skimming and Navigating How-To Video}
When following how-to videos, people engage in complex navigation behaviors -- rewatching certain parts to identify details and skipping irrelevant parts~\cite{chang2021rubyslippers}. To facilitate navigation, prior research explored segmenting based on tool usage~\cite{truong2021automatic, grossman2010chronicle}, intermediate results~\cite{kim2014crowdsourcing}, and other learners' interaction data~\cite{yang2022softvideo}. These navigation units are often visualized as multi-modal summaries with keyframes and transcript snippets ~\cite{truong2021automatic, pavel2014video, chang2021rubyslippers}, or graph-based visualization~\cite{liu2018conceptscape, kim2023surch}. Researchers have also explored how to contextually present these how-to video segments based on tool context~\cite{fraser2019replay, banovic2012waken, nguyen2015making} and physical space~\cite{thoravi2019tutorivr, sadprasid2024improving}. 
Compared to manually scrubbing through the video timeline, these navigational units make it easier to jump to meaningful moments. However, they often rely on visual interaction, which can be difficult when viewers' eyes and hands are occupied with the physical task. Thus, researchers proposed voice-based video navigation~\cite{chang2021rubyslippers, truong2021automatic, zhao2022rewind}.
Building on this prior work, we segment videos into structured steps and support voice-based control. While prior systems take explicit user commands for navigation, we detect users' step completion and proactively suggests moving to the next step -- enabling mixed-initiative guidance for time-sensitive tasks like cooking.

    

\subsection{Physical Task Assistance}
People following procedural tasks like cooking often rely on voice assistants (\textit{e.g.,} Amazon Alexa, Google Home) to read out instructions~\cite{ju2001counteractive}, answer questions about ingredients, or set timers~\cite{hwang2023rewriting, jaber2024cooking}. With advances in speech recognition and natural language understanding, these assistants now go beyond fixed commands and can infer user intent from free-form speech.
Beyond voice interactions, wearable assistants can enable richer context understanding with visual, audio, and motion sensors~\cite{chi2007enabling}. For example, head-mounted AR systems can be used to disambiguate user questions through visual context~\cite{lee2024gazepointar} and to convey the affordances of kitchen tools to low-vision users~\cite{lee2024cookar}. To support task assistance through a more common consumer device, Arakawa et al. developed smart watch systems that deliver context-aware support during physical tasks~\cite{arakawa2024prismObserver, arakawa2024prismQA}.

In everyday life, BLV people frequently use crowd-powered tools (\textit{e.g.,} Aira, Be My Eyes) and AI-powered applications (\textit{e.g.,} ChatGPT, Seeing AI) to identify objects~\cite{bigham2010vizwiz, huh2024long} or receive feedback on their work~\cite{huh2024designchecker, zhang2023a11yboard}. HCI researchers have also developed task-specific assistants to guide users in home exercises~\cite{rector2013eyes} and makeup~\cite{ishikiriyama2017interactive}. In this work, we implement a task assistant for guiding hands-on tasks (\textit{e.g.,} cooking) using smartglasses. As BLV individuals increasingly use smartglasses in their daily routines~\cite{tran2025wearable}, our system leverages the multimodal sensing capabilities of this platform to provide context-aware, hands-free guidance. 

\subsection{Computer Vision for Accessibility}
The capability to assist with physical tasks is closely linked to computer vision innovations. How-to videos have served as a valuable data source~\cite{miech2019howto100m} for training video understanding models capable of tasks like action recognition~\cite{ashutosh2023video,zhou2023procedure}, step localization~\cite{xue2024learning,dvornik2023stepformer}, skill assessment~\cite{ashutosh2024expertaf} and video question answering~\cite{zhong2022video}. The recent advent of large language models has catalyzed the development of vision-language Models (VLMs)~\cite{li2024llava,wang2024qwen2,achiam2023gpt,team2023gemini, astra}, which offer a unified natural language interface for diverse vision tasks and present a promising avenue for analyzing user videos real time. Despite their potential, current VLMs face substantial hurdles when used for physical task guidance, especially for BLV users. One primary issue is hallucination~\cite{bai2024hallucination}, where models produce information not grounded in the visual evidence. Also, these models often generate guidance presupposing visual capabilities, leading to suggestions unsuitable for BLV users~\cite{bigham2010vizwiz,song2024ego4d}. Further, existing VLMs often lack fine-grained analysis capabilities~\cite{xue2024progress}, struggling to monitor detailed execution progress within tasks, which is vital for step-by-step assistance. 
In this work, we carefully design task support grounded in BLV people's needs in real-world scenarios and propose a pipeline to generate accessible task guidance.



\section{Expert Observational Study}\label{sec:formative_study}

To reveal accessibility challenges in cooking, prior work conducted interviews with BLV individuals and cooking instructors~\cite{li2024recipe}, and observed how BLV people cook~\cite{li2021non, li2024contextual}. Building on these insights, we conducted an observational study where we monitored how vision rehabilitation therapists (VRTs) deliver real-time remote guidance to BLV individuals in following how-to videos. 
We observed VRTs as they are highly trained in teaching new skills and providing structured guidance to BLVs, establishing a strong benchmark for AI-driven assistance. 
While our study is not a contextual inquiry~\cite{beyer1999contextual} -- as BLV participants typically learn to cook in person or receive only intermittent remote assistance via phone -- we incorporated contextual elements into our study design by observing participants in their own kitchens, using their own tools, wearing their personal smart glasses, and working on recipes they selected.
This unique setting allows us to understand how accessibility experts provide adaptive, context-specific instructions and feedback and identify real challenges associated with remote assistance. 

 


\subsection{Method}~\label{sec:formative_method}
We recruited 3 VRTs (VRT1-VRT3) who regularly teach cooking skills to BLV students and 3 BLV participants (P1-P3) with cooking experience using mailing lists (Tables~\ref{tab:vrt_participants}-\ref{tab:blv_participants}). The study was conducted remotely via Zoom in 2 sessions (each 1.5 hours, 50 USD)\footnote{Approved by the institution’s Institutional Review Board (IRB).}. 
In the first session, each VRT was invited for a 1.5-hour study to create accessible cooking instructions from how-to videos. We provided each VRT with a cooking video — Chocolate chip cookies~\cite{cookies_video}, Fruit pavlova~\cite{pavlova_video}, or Eggs benedict~\cite{benedict_video} — and its transcript. 
We selected recipes that BLV participants were interested in learning, had no prior experience with, and could finish during the study. VRTs were asked to rewrite the video instructions to better suit the needs of the BLVs (results in Supplementary Materials). We also conducted a semi-structured interview to understand VRTs' current practices for teaching cooking and their strategies and challenges in making the cooking tutorial video more accessible.

In the second session, we observed VRTs as they remotely guided BLVs using these adapted materials. BLV participants wore their own Ray-Ban Meta smart glasses and joined the video call from their own kitchens to allow VRTs to monitor their activities in real time. We chose this setup instead of in-lab study as cooking processes can be highly contextual, depending on individual tools and kitchen layouts~\cite{li2024contextual}, and BLVs often use adapted kitchen tools~\cite{li2021non}. All participants had more than 5 years of cooking experience and were experienced in using heat-based appliances like stoves and ovens.
After the cooking task, we conducted separate semi-structured interviews with BLV and VRT participants. We asked BLV participants how the experience compared to their current practices, any challenges in interpreting and following instructions, and desired support in future cooking tasks. We asked VRT participants about their strategies for providing accessible instructions and feedback, along with any challenges they encountered.

We conducted a conversational analysis~\cite{goodwin1990conversation} of video recordings and annotated transcripts, attributing each utterance to either the BLV or VRT participant and segmenting at the sentence level. We applied four labels to structure our analysis: \textit{instruction}, \textit{supplementary} (tips or workarounds), \textit{progress description} (updates on the BLV participant’s task state), and \textit{question}. Responses were categorized into these labels based on content. See Supplementary Material for the full coding process and data.


\begin{figure}[t]
  \centering
  \includegraphics[width=\columnwidth]{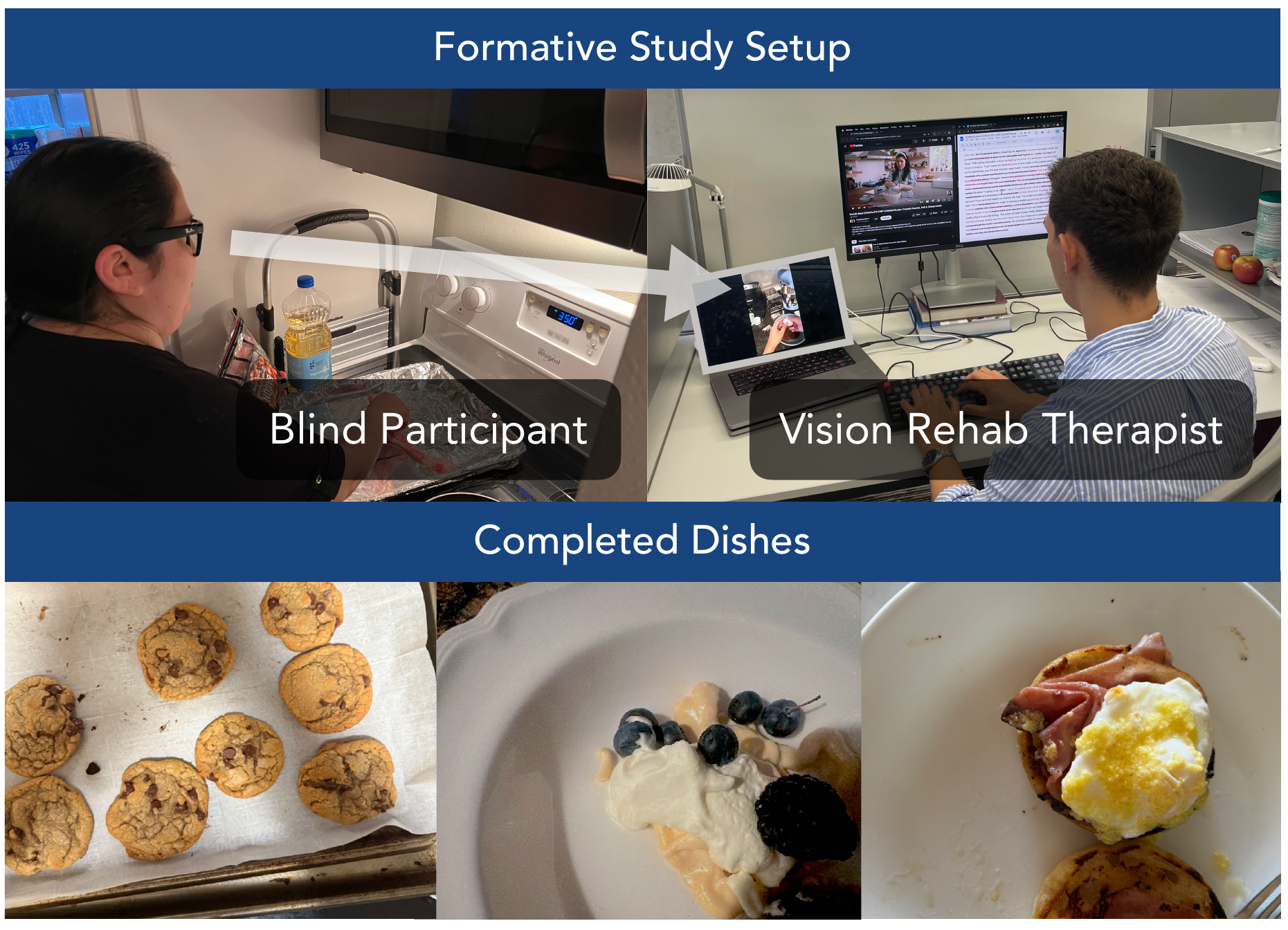}
  \caption{We observed how VRTs deliver real-time remote guidance to BLV individuals in following how-to videos while observing their progress through the smart glasses' video stream. Participants were assigned to how-to videos on making chocolate chip cookies (P1 \& VRT1), fruit pavlova (P2 \& VRT2), and eggs benedict (P3 \& VRT3).}\label{fig:formative}
\end{figure}

\begin{figure*}
  \includegraphics[width=\textwidth]{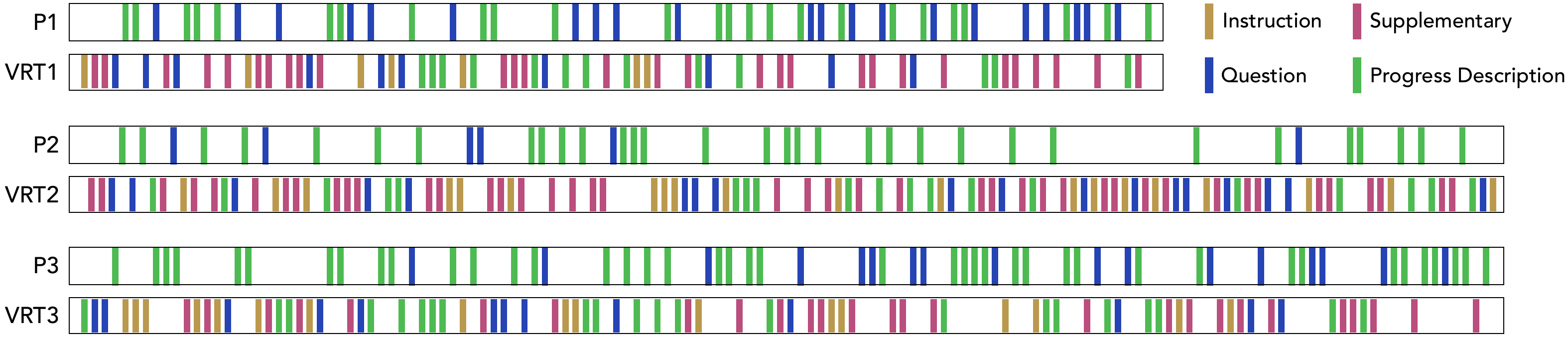}
  \caption{Timelines of utterances annotation of observational study sessions. \camready{Colors denote utterance types and the x-axis plots interaction order.} VRT2 provided more proactive supplementary information and progress descriptions to novice BLV2.}
  \label{fig:formative_annotation}
\end{figure*}

\subsection{Current Practices}
\ipstart{Accessing Online Recipes}
To learn how to cook a new recipe, all BLV participants used online resources (\textit{e.g.,} text recipes, how-to videos). 
Echoing the findings of prior work~\cite{liu2021makes, li2024recipe}, participants searched for videos with rich narration and less on-screen text. P2 strategically searched for popular videos as they tend to be more descriptive and P3 considered videos from creators with similar kitchens and tools. 
P1 mentioned~\textit{``I don't like blind-targeted resources, they're limited in variety and focus only on simplifying the steps rather than achieving high-quality.''}
Even with narrated videos, there were navigation challenges. P3 shared ~\textit{``These videos tend to have very long sentences and give multiple instructions at once, so I need to rewind back to break it down.''} Also, instructions were often high-level and did not provide sufficient information to correctly follow the step. Thus, P2 mentioned having a lot of follow-up questions when following how-to videos. ~\textit{``She [The video creator] would tell me to cut the carrots, but should I dice it? Cut into cubes? How many inches? Are the carrots peeled in the video?''}

\ipstart{Human and AI assistance}
Participants also learned how to cook from VRTs (P1-P3) and families and friends who cook (P1-P3). P3 noted ~\textit{``Sometimes it's hard to find the right person because they should know how to cook and convey ideas non-visually.''}
Compared to learning from online resources, cooking with sighted people was easier as they could provide a lot of workarounds and give feedback on the cooking progress. P3 mentioned ~\textit{``People who lost their vision recently may have a fear of heat or spilling things. VRTs can tweak the recipe to minimize this stress.''} 
Also, P2 asked sighted people to compare the visuals in a how-to video and her dish to assess her progress. 
P1 and P2 also used visual interpreter services (\textit{e.g.,} Aira, BeMyEyes) but noted that interpreters often lack cooking expertise.

Participants occasionally used visual understanding AI (\textit{e.g.,} ChatGPT) for identifying spices (P1), learning how to use a new tool (P3), and checking the doneness of the food (P1, P2). However, not all information from AI was relevant (\textit{e.g.}, kitchen decorations - P1).
P3 who doesn't use AI for feedback explained ~\textit{``I can assess my progress with other senses like smell, sound, or even by touching. These skills are more reliable and make me independent.''}
VRT participants mostly provided cooking lessons in-person for groups. 
They all highlighted the importance of adapting the lessons to individual students' vision, cooking experience, and kitchen setup. 
While VRT2 has tried providing remote cooking guidance through a video call, it was hard for students to hold their phones at the right angle.

\subsection{Study Findings}~\label{study_findings}
From observing how VRTs rewrite how-to video instructions (Session 1) and provide remote task assistance to BLV individuals (Session 2), we distill key takeaways that inform the design of Vid2Coach.


\ipstart{Session 1 - Rewriting Recipes}
On average, VRTs made 8 (SD=5.57) modifications to the original video transcript to improve the description of visual content. How-to videos often used unclear visual references in their narrations (\textit{e.g., ``First, I will pour \textbf{this} over \textbf{here}.''}) without explicitly naming the tools and ingredients. To address this, all VRTs replaced vague references with specific object names. Even when an object is already specified (\textit{e.g.,} a pan), VRT2 further detailed its attributes (\textit{e.g.,} a medium size, stainless pan). Additionally, VRT1 provided descriptions of \textit{how} an action is being performed. When the chef in the video instructed to add one cup of brown sugar, VRT1 clarified that it was loosely measured rather than packed. VRT1 noted ~\textit{``These small details can make the difference in the outcome, so making sure I don't miss them.''}
Some actions were only demonstrated visually without verbal mention, making them completely inaccessible to BLV followers. VRTs addressed this gap by explicitly describing these omitted steps. 

Since BLV users rely entirely on verbal descriptions, it was essential to provide clear and structured instructions. VRTs organized the content by listing ingredients and tools before each step. They also broke long sentences into single actions. Additionally, they removed unnecessary chatter and reordered steps for efficiency. VRT2 mentioned ~\textit{``They [BLV] are not watching the visuals and have to process everything by listening so it's harder.''}
\begin{center}
\begin{minipage}{\linewidth}
\begin{mdframed}[
    linewidth=0.8pt,
    backgroundcolor=gray!10,
    linecolor=black,
    innerrightmargin=6pt,
    innerleftmargin=6pt,
    innertopmargin=4pt,
    innerbottommargin=4pt
]
\textbf{Takeaway:} Provide detailed instructions based on both the narration and visuals from the video. Structure rich information into clear, concise segments to support easier understanding.
\end{mdframed}
\end{minipage}
\end{center}
 VRTs did not simply relay video instructions but made 15 (SD=7) edits to provide tips and workarounds to make the steps easier to follow. For instance, VRT2 modified the egg-cracking step to first crack the egg into a separate bowl so that BLV participants can use their hands to check for shell fragments before adding it to the mixture. In addition to workarounds, VRTs provided safety reminders (\textit{e.g., ``Be careful when taking the cookies out of the oven. Don’t forget the oven mitts,''} -- VRT1) and practical tips for independent cooking (\textit{e.g., ``You can add tactile stickers to your mixer's speed control for next time when you're cooking alone.''} -- VRT2). 

However, VRTs found it difficult to remember accessible cooking techniques. VRT1 explained, \textit{``Cooking is just one of many skills we [VRTs] teach -- navigation, transportation, technology, and more. We can’t remember every accessible cooking tip, so we often have to search online before teaching a new recipe.''} Also, there was a lack of a centralized resource on accessible cooking practices. VRTs thus spent time and effort gathering information from multiple sources.

\begin{center}
\begin{minipage}{\linewidth}
\begin{mdframed}[
    linewidth=0.8pt,
    backgroundcolor=gray!10,
    linecolor=black,
    innerrightmargin=6pt,
    innerleftmargin=6pt,
    innertopmargin=4pt,
    innerbottommargin=4pt
]
\textbf{Takeaway:} Supplement instructions with tips and accessible workarounds for both immediate guidance and building reusable strategies for future tasks.
\end{mdframed}
\end{minipage}
\end{center}

\ipstart{Session 2 - Realtime Task Assistance}
All BLV participants expressed excitement in getting real-time remote cooking guidance for following how-to videos and getting feedback. During the study, BLV and VRT participants engaged in rich, collaborative dialogue (Figure~\ref{fig:formative_annotation}). All VRTs offered visual descriptions both proactively and in response to BLV questions. When participants were engaged in independent actions (\textit{e.g.,} washing hands, finding the glass bowl), VRTs paused their guidance and waited. When BLV participants were performing actions, VRTs provided visual description of their progress (\textit{e.g.,} \textit{``You seem to be halfway done, but I still see some flour residue.''} -- VRT1). Compliments were also frequent (\textit{``Perfect, you've made a beautiful hollandaise.'' -- VRT2}) which they used to support BLV participants' confidence as VRT2 explained -- \textit{``I do so because it can help with their self-esteem. This should be a fun and enjoyable process.''} When BLV participants seemed to be complete with a step, VRTs often described it as done and moved on to the next instruction. However, from post-task interviews, we learned that BLV participants sometimes preferred more specific visual cues for gauging completeness. For example, P1 mentioned ~\textit{``I wanted her [VRT1] to describe what makes it look complete by sight rather than just saying it's done. So that I can be sure.''} 

Despite the benefits, remote assistance introduced several challenges. To provide visual feedback, VRTs sometimes had to ask BLV participants to adjust the glasses’ camera angle, as the view was often too narrow or not focused on their hands~\cite{vazquez2014assisted}. In addition, due to unstable video quality with streaming setting, some BLV participant mistakes were unnoticed. 

\begin{center}
\begin{minipage}{\linewidth}
\begin{mdframed}[
    linewidth=0.8pt,
    backgroundcolor=gray!10,
    linecolor=black,
    innerrightmargin=6pt,
    innerleftmargin=6pt,
    innertopmargin=4pt,
    innerbottommargin=4pt
]
\textbf{Takeaway:} Provide proactive visual feedback on user progress and success, and prompt users when adjustments are needed to keep relevant actions in view.
\end{mdframed}
\end{minipage}
\end{center}
As how-to videos often prioritize visual descriptions of the cooking progress~\cite{li2024recipe, bilyk2009food}, VRTs often translated visual descriptors into other sensory cues to help BLV individuals assess progress independently. For example, instead of saying ~\textit{``The butter has started to melt.''}, VRT1 described ~\textit{``You should feel them soften with the spatula, or hear some sizzling sound.''}
Also, with remote assistance setting, VRTs found it difficult to evaluate certain steps with subtle visual cues and asked BLV participants how they feel, hear, or smell (\textit{e.g.,} VRT2 asking P2 if the meringue felt stiff enough to hold its shape).

However, we observed a tension between encouraging non-visual assessment and providing proactive visual feedback. VRT2 intentionally encouraged P2 to rely on touch or smell to assess progress independently, explaining, ~\textit{``I try not to give much visual feedback. I asked her to check for herself because I wanted her to be independent. What if AIRA [Visual interpreter service] isn’t available? What if the internet isn’t working? High-tech solutions are great, but low-tech methods are more reliable—especially when using a phone is difficult, like when you're cooking or in the shower.''} However, from the post-task interview, P2 mentioned that she wanted more frequent visual confirmation during the task for reassurance.


\begin{center}
\begin{minipage}{\linewidth}
\begin{mdframed}[
    linewidth=0.8pt,
    backgroundcolor=gray!10,
    linecolor=black,
    innerrightmargin=6pt,
    innerleftmargin=6pt,
    innertopmargin=4pt,
    innerbottommargin=4pt
]
\textbf{Takeaway:} Elicit users to leverage non-visual sensory cues to evaluate the progress, to build confidence and support independent task completion.
\end{mdframed}
\end{minipage}
\end{center}

Despite receiving detailed descriptions, tips, and feedback, all BLV participants still asked questions throughout the task. Participants asked for repetition or clarification of instructions. P1 often asked VRT1 to read upcoming steps so that she could adapt the sequence or combine steps in a way that fits her workflow. BLV participants also asked for visual feedback when VRTs were not proactively offering it. A significant portion of the questions focused on verifying their actions against the how-to video demonstration -- whether they were using the correct tool, ingredient, or amount. Answering these questions often required VRTs to revisit their rewritten instructions or visually compare BLV participants' actions with the original video content, highlighting the importance of situating responses within the specific task context.

\begin{center}
\begin{minipage}{\linewidth}
\begin{mdframed}[
    linewidth=0.8pt,
    backgroundcolor=gray!10,
    linecolor=black,
    innerrightmargin=6pt,
    innerleftmargin=6pt,
    innertopmargin=4pt,
    innerbottommargin=4pt
]
\textbf{Takeaway:} Address users’ diverse questions by grounding responses in both the user’s current task progress and the how-to video knowledge, enabling flexible task execution.
\end{mdframed}
\end{minipage}
\end{center}

All VRTs asked BLV participants about their vision level, cooking experiences, and available tools to tailor the guidance.
For instance, VRT2 provided more frequent instructions and supplementary information to P2, who was new to baking (Figure~\ref{fig:formative_annotation}). Similarly, VRT3 first asked P3 whether he had ever separated egg yolks from whites to determine how much detail was needed.
VRTs also adapted instructions to the user’s environment by referencing details visible in the egocentric video, giving spatial cues like \textit{``Grab that flour on your right.''} or using clock-face directions such as \textit{``For medium heat, set it to 5 o'clock.''} They also modified instructions based on BLV participants' kitchen set up. For instance, when guiding P1 to add butter to a bowl, VRT1 searched online to convert the cup measurement into weight, explaining, \textit{``She [P1] had a talking scale, so I changed the instruction to use it instead of measuring with cups since that’s easier.''} 





\begin{center}
\begin{minipage}{\linewidth}
\begin{mdframed}[
    linewidth=0.8pt,
    backgroundcolor=gray!10,
    linecolor=black,
    innerrightmargin=6pt,
    innerleftmargin=6pt,
    innertopmargin=4pt,
    innerbottommargin=4pt
]
\textbf{Takeaway:} Adapt guidance to user skills, preferences, and context by adjusting the level of detail, rephrasing based on familiarity, and referencing available tools or spatial layout.
\end{mdframed}
\end{minipage}
\end{center}

\noindent Based on our observations, we distill 5 design goals for a system that provides video-based task guidance for BLV people. While our formative study focused on cooking task assistance, we believe these insights can be applied to support similar hands-on activities, such as crafting, home workouts, and DIY repairs.

\begin{itemize}
    \item[\textbf{D1.}] {Provide instructions based on both narration and visual demonstrations of how-to videos} 
    \item[\textbf{D2.}] {Supplement instructions with accessible tips and workarounds}
    \item[\textbf{D3.}] {Provide proactive visual feedback on user progress}
    \item[\textbf{D4.}] {Encourage users to leverage non-visual sensory cues to evaluate the progress}
    \item[\textbf{D5.}] Address users' diverse questions with responses grounded in users' task progress and the how-to video knowledge
    \item[\textbf{D6.}] {Adapt instructions and feedback to user preference, skills, and context}
    
\end{itemize}




\section{Vid2Coach}
We present Vid2Coach, a system that transforms how-to videos into
a task assistant on smartglasses (Figure~\ref{fig:teaser}).
Grounded in how-to videos and accessible resources, Vid2Coach provides detailed instructions, accessible workarounds, and real-time feedback to support BLV people in following a procedural task. 

\subsection{User Scenario}
To illustrate the use of \sysname{}, we follow Ash, a blind student who often uses Vid2Coach to follow recipe videos. Recently, Ash enjoyed Eggs Benedict at a restaurant and wants to make it himself. Among many how-to videos, Ash chooses Mary Berry’s BBC-featured Eggs Benedcit recipe which is popular and well-reviewed. After entering the YouTube link into Vid2Coach's web app, the system generates a high-level summary of the steps, ingredients, and tools needed, which he can access with his screen reader.

Once he has gathered the necessary ingredients and tools, Ash puts on his smart glasses and connects them to Vid2Coach. His accessibility preferences are already saved in the system, including his vision level and assistive kitchen tools (\textit{e.g.,} talking thermometer, bump dots), so he can start cooking right away. Vid2Coach reads the first instruction aloud: ~\textit{``Heat the pan until it’s hot.''} with the demonstration details of how Mary is performing the task ~\textit{``The person in the video uses a medium to large black skillet without any oil.''} \textbf{(D1)} so Ash finds a similar one and begins heating it on the stove. 
With the next action -- \textit{``Add 8 strips of bacon''}, Ash asks \textit{``Do I need to add any oil?''} and the system responds \textit{``According to the video, no oil is needed because the bacon will render its own fat.''} \textbf{(D5)}

While cooking the bacon, Vid2Coach proactively describes the bacon's color at each stage: when it has a pink hue, then when it turns brown, and finally when it reaches dark brown. Vid2Coach then says ~\textit{``You seem to be done because the bacon looks dark brown and crispy. You can also check by feeling its crispiness with a fork or listening to the sizzling sound. Would you like to move on?''} \textbf{(D2-D3)} He tests the texture with a fork, confirms its crunchiness, and turns off the heat just before it burns. 
While Ash washes his hands, Vid2Coach pauses its description and waits.

Next, Vid2Coach guides Ash through the separation of eggs for the hollandaise sauce. It describes how the chef in the video performs the action using the traditional two-shell method, but Ash is not confident in following it and asks for an easier workaround. Vid2Coach suggests a technique from a blind cook's video: cracking the egg into a bowl and using fingers to gently lift the yolk while letting the whites slip through. \textbf{(D4)} Ash successfully follows the workaround. Later, while seasoning the sauce, he realizes his new salt and pepper containers are not braille-labeled yet, and asks Vid2Coach  \textit{``Which one is which?''} Since his glasses were not facing the containers, Vid2Coach prompts him to adjust his view. Once he does, it identifies: \textit{``Salt is on your right, and pepper is on your left.''}
When Ash asks how to prevent the sauce from turning into scrambled eggs, Vid2Coach recalls the tools in Ash's kitchen and replies, ~\textit{``You can use your talking thermometer to check the temperature -- it should be around 120°F according to the video.''} \textbf{(D5-D6)}

When assembling the benedict, Ash forgets about the correct order and asks Vid2Coach to remind the order, and it responds: ~\textit{``Place the toasted bread first, then the poached egg, followed by the hollandaise sauce, and finish with 2 pieces of bacon on top.''} 
After presenting them, Ash says he is finished, but Vid2Coach reminds him: ~\textit{``You forgot to add bacons to one of the toasts.''} \textbf{(D2)}
He fixes the mistake and serves the dish with confidence.

\section{Pipeline}
Vid2Coach's pipeline generates step instructions from a how-to video (\S\ref{sec:generate_steps}), supplements tips and workarounds from accessibility task resources (\S\ref{sec:supplementing_strategies}), monitors users' progress to provide proactive feedback (\S\ref{sec:pipeline_monitoring}), and answer users' questions (\S\ref{sec:intent_elicitation}).

\subsection{Generating Step Instructions}\label{sec:generate_steps}
\begin{figure}[t]
  \centering
  \includegraphics[width=\columnwidth]{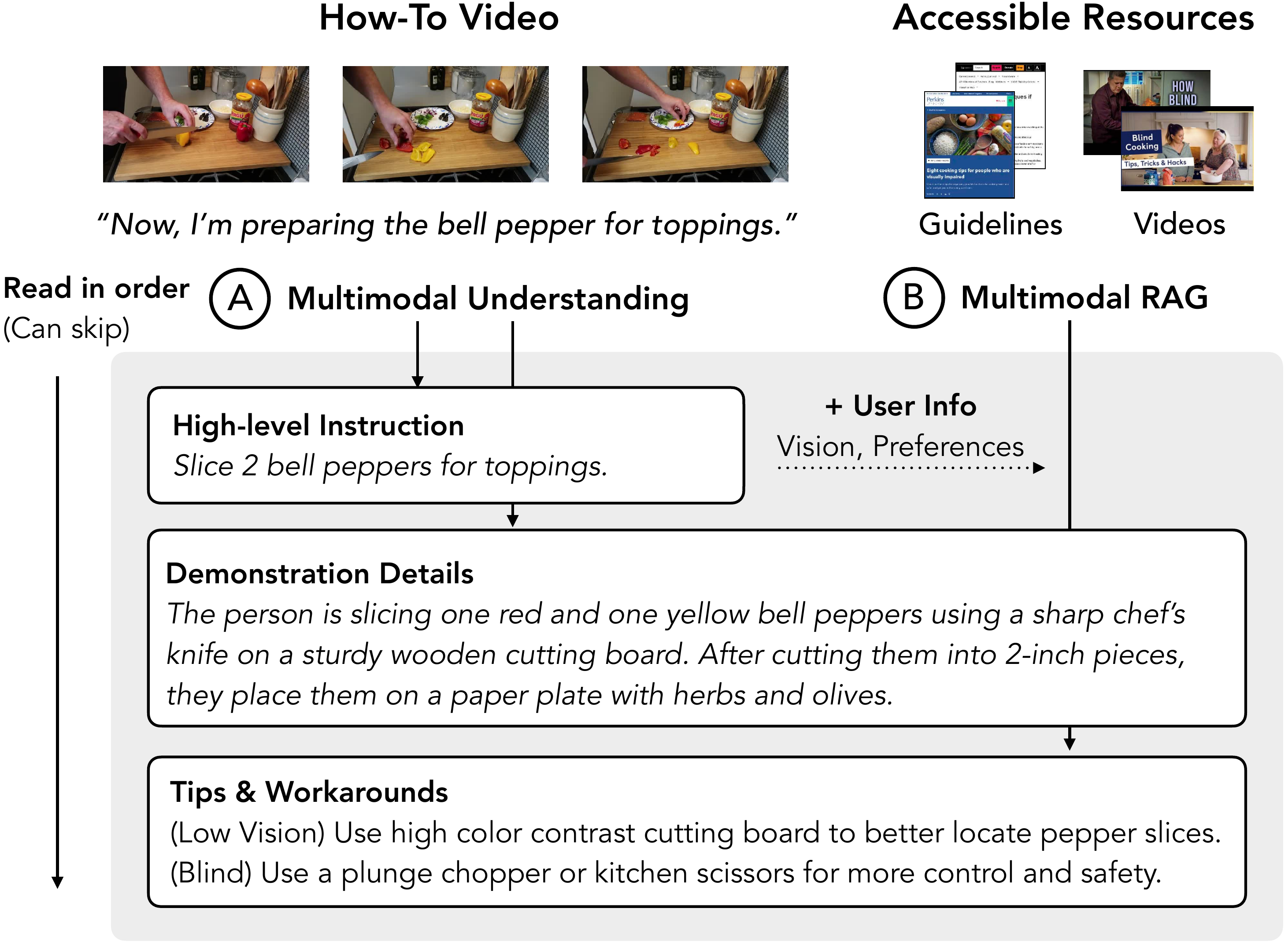}
  \caption{Vid2Coach generates step instructions from a how-to video with multi-modal understanding of narration and frames (A), and supplements tips and workarounds from accessibility task resources using RAG (B).\vspace{-30pt}
  }\label{fig:pipeline1}
\end{figure}

Following findings from our observational study, Vid2Coach's pipeline generates structured instruction with detailed demonstration descriptions (Figure~\ref{fig:pipeline1}A).

\ipstart{Generating Step-Action Hierarchy from Narration}
Vid2Coach first transcribes the video narration using Whisper~\cite{whisper}, then segments the transcript into sentences using punctuation and word-level timestamps. Prior research has shown that how-to videos often contain narration beyond instructions such as tips, warnings, and self-promotion~\cite{yang2023beyond}. 
To only retain task-relevant narration, we use an LLM (GPT-4o~\cite{achiam2023gpt}) to classify each sentence into one of eight information roles derived from Yang et al.'s taxonomy~\cite{yang2023beyond} and filter greeting, conclusion, and miscellaneous role sentences.
Then, we use an LLM (GPT-4o~\cite{achiam2023gpt}) to group relevant sentences into high-level steps. Prior approaches segmented videos based on scene transition~\cite{huh2023avscript}, object appearance~\cite{chang2021rubyslippers}, or intermediate outcomes~\cite{yang2025videomix}. 
Inspired by Truong et al.~\cite{truong2021automatic} using hierarchical segmentation, we generate \textbf{high-level steps} that help users understand the goal (\textit{e.g., Prepare hollandaise sauce}), accompanied by \textbf{atomic actions} that helps users to easily follow and assess (\textit{e.g., Separate 3 egg yolks from the whites}). 
Sentences with multiple actions (\textit{e.g., ``Add 1.5 cups of melted butter and 1 tablespoon of vinegar, then mix until fully incorporated''}) are split into atomic actions centered around a single verb. For each step, Vid2Coach extracts relevant tools and ingredients to help users prepare the step, and saves sentences labeled as non-instructional (\textit{e.g.,} tips, warnings, or explanations) to provide on demand. (See all prompt details in \S\ref{sec:pipeline_prompts})

\ipstart{Identifying Task-Relevant Frames}
How-to videos often include visuals that are not directly relevant to the instructional steps, such as a talking-head view of the presenter or B-roll footage of the dish’s origin. Including these in the description pipeline can lead to irrelevant or distracting descriptions, so we extract only task-relevant frames. We sample video frames at 1 frame per second within a ±15-second window around the start and end of each action's transcript segment.
This window captures visual demonstrations that may occur slightly before or after the spoken instruction and reduces hallucination by filtering out more distant segments with similar objects or tools.
Then, we compute cross-modal similarity between each frame and action description using CLIP~\cite{radford2021learning} and only keep those above a dynamic threshold (ranging from 0.27-0.30). This threshold is adjusted per action based on frame density: increasing when many frames exceed the threshold and decreasing when few do.
For instance, in videos where demonstrations are filmed from a wide-angle shot, relevant actions may appear smaller and result in lower scores despite being relevant. In such cases, the threshold is lowered to retain useful frames.

\ipstart{Generating Task-Relevant Descriptions}
Finally, we combine the step information from narration and the high-scoring frames and use a VLM~\cite{achiam2023gpt} to generate demonstration details for each action. When prompted naively, VLMs often include details such as the presenter’s appearance, decorative backgrounds, or camera angles. To guide the model toward task-relevant content, we follow Huh et al.~\cite{huh2023genassist} and use a targeted prompting strategy based on questions derived from our formative study: (1) \textit{What is the demonstrated action?} (2) \textit{Which ingredients are used, how do they look, and how much is being used?} (3) \textit{Which tools are used, and how do they look?} (4) \textit{How is the action performed?} and (5) \textit{Are there any tips for performing this action evident from the images?} 

\subsection{Supplementing Accessible Strategies}\label{sec:supplementing_strategies}
Informed by VRTs in our formative study who supplemented video instructions with safety warnings and accessible workarounds, we leverage grounded accessibility resources to extract relevant information and augment the description (Figure~\ref{fig:pipeline1}B). 

\ipstart{Data collection} 
We curated two task-specific datasets: (1) cooking and (2) arts \& crafts. The cooking dataset includes 100 non-visual cooking videos created by BLV individuals and 100 publicly available accessibility resources (\textit{e.g.,} blog posts, community guides, instructional articles). The arts \& crafts dataset includes 30 videos by BLV creators and 50 relevant accessibility resources. Following prior work on surfacing accessibility content~\cite{li2021non, li2022feels}, we used a combination of vision-related (\textit{e.g.,} ``blind,'' ``low vision'') and task-related (\textit{e.g.,} ``kitchen'', ``crafts'') search keywords. We include the list of URLs in the supplemental materials and will release the dataset.

\ipstart{Retrieval Augmented Generation}
General-purpose AI suggestions often assume dominant user groups, and may generate irrelevant suggestions that make BLV users feel misfit (\textit{e.g., Check if the butter looks golden brown})~\cite{alharbi2024misfitting}. Even when prompted to provide accessible tips, these models reflect biased assumptions and suggest that users seek help from sighted people or offer overly generic advice without practical value -- reminding them to be careful while cooking.
To address these limitations, we use a multi-modal retrieval augmented generation (RAG) approach~\cite{lewis2020retrieval} to ground AI suggestions in real-world, disability-aware resources. From the curated dataset, our pipeline first extracts text chunks to generate text embeddings, and identifies images and describes them with a VLM (Gemini 1.5 pro) to generate embeddings. At inference time, we input queries to generate step-specific workarounds, and the query also incorporates user-specific context to further personalize suggestions (\textit{e.g.,} vision level, cooking experience, available tools and kitchen setup) (see \S~\ref{tab:tip_workaround_prompt} for details). We compute the cosine similarity between the query's text embedding and the stored embeddings and retrieve top-3 text chunks to generate answers with the context.
\camready{It then aggregates these text chunks into a tip, surfacing multiple alternatives when appropriate (Figure~\ref{fig:teaser}.2).}

\subsection{Progress Monitoring}\label{sec:pipeline_monitoring}
\begin{figure}[t]
  \centering
  \includegraphics[width=\columnwidth]{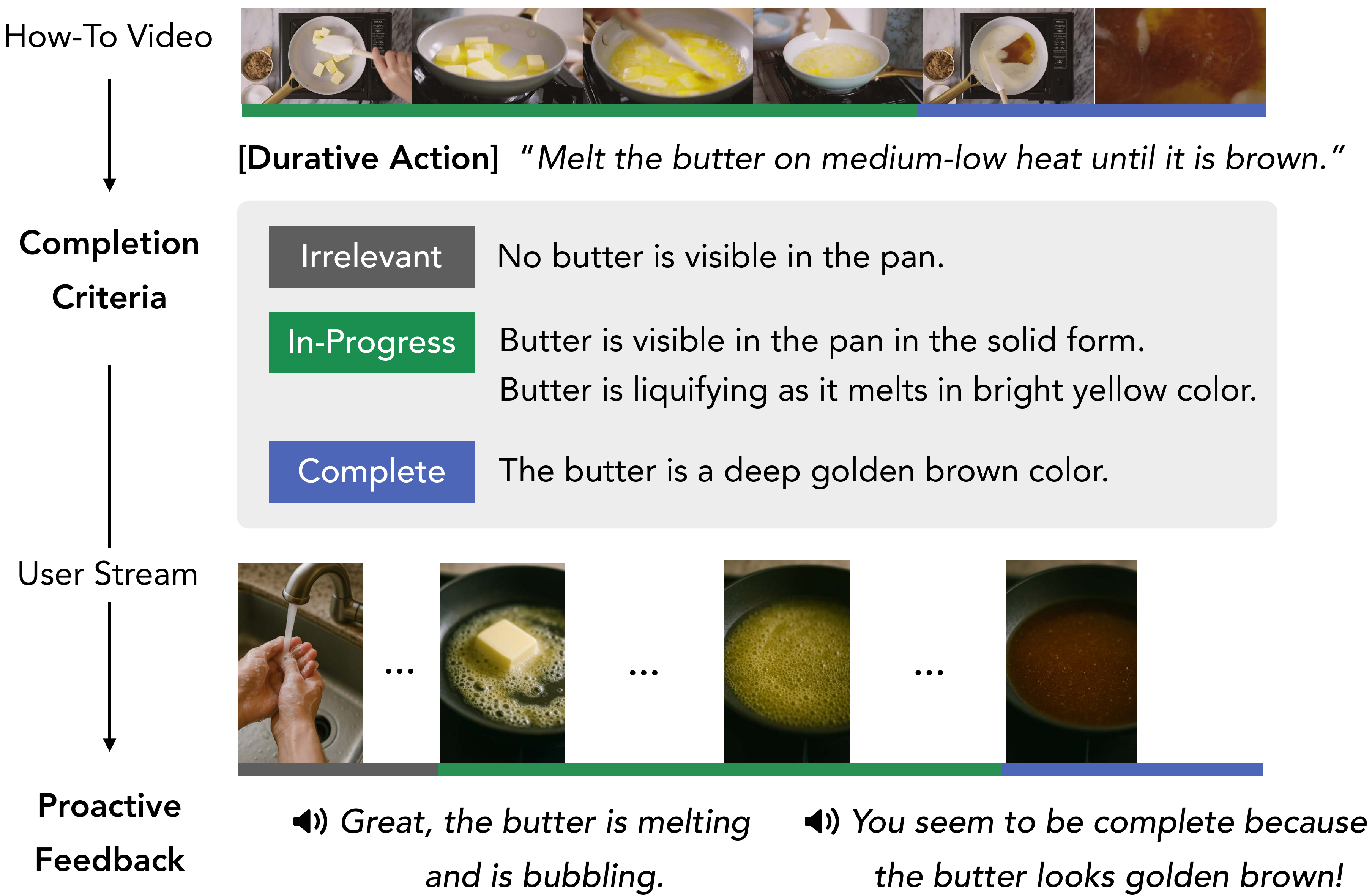}
  \caption{From the how-to video, Vid2Coach generates criteria for classifying user status into irrelevant, in-progress, and complete. As user performs the task, Vid2Coach monitors the progress and provide realtime feedback.}\label{fig:pipeline_monitoring}
\end{figure}

From the how-to video, our pipeline generates a set of criteria (irrelevant, in-progerss, complete, mistake) for each action and uses these to monitor the user’s progress (Figure~\ref{fig:pipeline_monitoring}). These criteria are derived using a VLM~\cite{achiam2023gpt} (details in \S\ref{sec:pipeline_prompts}). 
Instead of directly comparing frames from the how-to and user streams, we use abstracted criteria that are more robust to variations in tools, occlusions, and user behavior. As users perform each action, Vid2Coach uses these criteria to adapt its feedback: it pauses during unrelated or impromptu actions (\textit{e.g.,} washing hands), provides progress updates while a step is in-progress, and proactively suggests moving on once the step appears complete. ~\camready{Prior work demonstrates that people adapt recipes by combining, skipping, or improvising steps~\cite{ashutosh2024detours, yang2025videomix, comber2013food}, a pattern mirrored by the VRTs we observed.
Thus, Vid2Coach lets users freely navigate the steps and evaluates each step independently rather than assuming a strict step order~\cite{li2025oscar}.}

To make this system effective in real-time streamed video, we propose a 3-way categorization of actions --
\textit{punctual, iterative,} or \textit{durative} -- based on its temporal characteristics. Punctual actions (\textit{e.g.,``add 1 cup of flour''}) happen quickly and may be missed in frame sampling or batched inference. For punctual actions, in-progress feedback is often untimely or unnecessary, so Vid2Coach only confirms completion. For iterative actions (\textit{e.g., ``place three scoops of cookie dough''}) that involve repetition, it counts visible repetitions for tracking progress. For durative actions (\textit{e.g., ``cook until golden brown''}) that involve gradual visual change, Vid2Coach offers real-time updates on its progression and completion. For actions where visual cues are minimal (\textit{e.g., ``wait for the mixture to cool down''}), Vid2Coach asks users to decide when to proceed. \camready{
Vid2Coach prevents false positive step progression by prompting a quick confirmation: -- \textit{``You seem to be complete because the butter looks golden brown. Would you like to move on to the next step?''}}

To balance the accuracy and speed of realtime feedback, we adopt a \textit{dual-model} approach. Inspired by WorldScribe~\cite{chang2024worldscribe}, which leveraged different models with varying latency and granularity, our setup combines 1) a batch model (Gemini 2.0) for deeper reasoning with 2) a lightweight streaming model (Gemini 2.0-Live) for immediate responsiveness. 
The batch VLM runs every 5 seconds and looks at the five most recent frames (sampled at 1 fps) to determine user status based on pre-generated completion criteria. Unlike batch models, streaming models begin generating output token-by-token before fully processing the input~\cite{xiao2023efficient}, making them less suited for complex reasoning but valuable for continuous monitoring.
Thus, we use the streaming VLM to provide real-time, low-latency descriptions of user actions when a durative action is in progress. Users can toggle this progress feedback on or off using a voice command. 
\camready{To avoid memory buildup and maintain inference consistency, Vid2Coach resets the model’s input context at each step, preventing cascading errors in the pipeline.}

\subsection{Intent Elicitation \& Question Answering} \label{sec:intent_elicitation}
In addition to automated feedback, Vid2Coach also responds to users' open-ended, voice-based queries. To ensure responsiveness, we use the streaming model's voice activity detection (VAD) to immediately interrupt any ongoing generation and prioritize the user's spoken query. Inspired by patterns observed in our formative study (\S\ref{sec:formative_study}), we implement a few-shot classification prompt that identifies the user's intent from both \textit{explicit} utterances (\textit{e.g., ``Does this look complete?''}) and \textit{implicit} expressions (\textit{e.g., ``I can’t tell which of these is sugar''}). We classify user utterances into the five intent types and trigger a different function call to handle each request:
\begin{itemize}
    \item \textbf{Navigation Commands}: Instructions such as \textit{``go back''}, \textit{``repeat that''}, or \textit{``next step''} make system navigate and repeat instructions.
    \item \textbf{Tips or Workarounds}: Requests such as \textit{``What’s an easier way to do this?'}' trigger the system to provide accessible suggestions from our RAG pipeline.
    \item \textbf{Progress Feedback Requests}: Questions like \textit{``Does this look ready?''} or \textit{``Is it done yet?''} are answered with batch model's analysis of user video stream using pre-generated completion criteria.
    \item \textbf{General Visual QA}: Questions like \textit{``Which of these is salt?''} or \textit{``What’s the expiration date?''} are answered with the batch model's analysis of user video stream.
    \item \textbf{Non-Visual Knowledge Queries}: Broader queries such as \textit{``How do I use a zester?''} or \textit{``What’s half of 3/4 cup?''} are answered with the streaming model with the recipe context.
\end{itemize}

For visual tasks where hallucinations are common and accuracy is important, our pipeline rely on the batch model’s reasoning capabilities, while the streaming model handles quicker, low-latency interactions such as navigation and general knowledge queries.

\subsection{Implementation}
We implemented our real-time assistant as a React-based web application that integrates with the Google Gemini Multimodal Live API~\cite{multimodal_live}, which supports low-latency bidirectional voice and video communication over WebSocket. 
Users’ progress captured with Meta glasses is streamed to a lab computer and fed into the React app for real-time monitoring and feedback. 

\section{Technical Evaluation}
We evaluate the robustness of our pipeline's ability to 1) generate detailed instructions from how-to videos and 2) monitor user progress through a real-time egocentric stream. 

\begin{figure*}
  \includegraphics[width=7in]{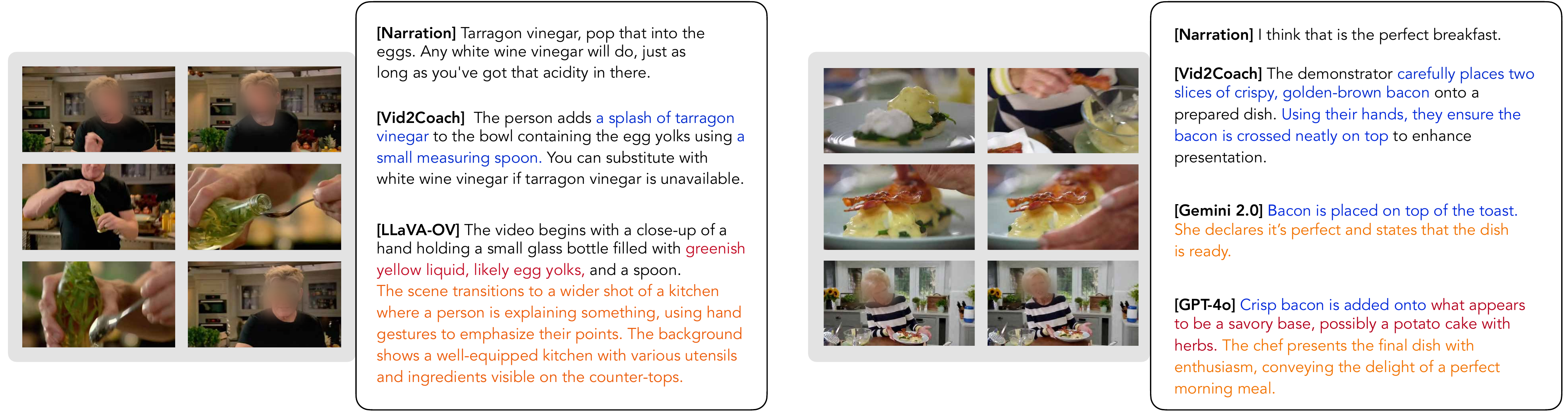}
  \caption{Qualitative comparisons of Vid2Coach descriptions with SOTA VLMs on 2 action sequences. These VLM descriptions often include hallucinations (red) and less task-relevant details (orange). Vid2Coach was able to capture new task-relevant details not covered in the narration (blue).}
  \label{fig:pipeline_qualitative}
\end{figure*}

\subsection{Generating Instructions}
\ipstart{Method}
We selected 10 cooking videos from YouTube. While the dataset is relatively small, this size allows for in-depth, manual annotation required for evaluation. Videos ranged from 4 to 8 minutes (SD=85 sec), and all had spoken narration. We selected a range of videos that varied in both dish type and narration density (\textit{e.g.,} minimal narration (V2) vs. detailed explanations (V9), see \S\ref{apndx:tech_eval_materials} for the list of videos).
We compared our pipeline against 3 state-of-the-art VLMs for video understanding tasks~\cite{patraucean2023perception, fu2024video} -- 2 proprietary VLMs (GPT-4o~\cite{achiam2023gpt}, Gemini 2.0~\cite{team2023gemini}) and 1 open-source VLM (LLaVA-OneVision~\cite{li2024llava}).
We used the same prompt for all models: \textit{``Describe the video in detail.''} For Gemini 2.0, which supports full-length video input, we provided the entire video directly. For GPT-4o and LLaVA-OneVision that require shorter inputs, we used the segmented steps from our own pipeline, and for each step, we tested two narration input conditions: 1) the extracted ~\textit{step name} and 2) the ~\textit{full narration} of the step. Comparing these two conditions allows us to evaluate how well the models leverage visual information alone versus how they perform when given more narrated context. For open-source models that have an input frame limit per call, we periodically subsample 32 frames from that step. 

To assess output quality, we annotated three aspects of each generated description: 1) \# of new task-relevant atomic facts that are not mentioned in the narration, 2) \# of missing task-relevant atomic facts, and 3) hallucinations. Because individual sentences can contain multiple discrete facts, we counted at the level of atomic facts~\cite{jing2023faithscore} to ensure fine-grained evaluation. We focused solely on task-relevant content (\textit{e.g.,} the state of ingredients or tools), excluding stylistic or background details (\textit{e.g.,} clothing or kitchen decoration). 
For the first two categories, we used an LLM (GPT-4~\cite{achiam2023gpt}) following prior work that shows its capability to extract and compare atomic facts from descriptions~\cite{jing2023faithscore, min2023factscore}. For hallucination detection, which often involves subtle visual grounding judgments, two researchers independently labeled hallucinated content and resolved disagreements through discussion.
    
\ipstart{Results}
As table~\ref{tab:pipeline_results1} summarizes, Vid2Coach outperformed baselines by identifying more task-relevant atomic facts and omitting fewer details present in the original narration. As shown in Figure~\ref{fig:pipeline_qualitative}, Vid2Coach surfaced concrete visual details that can support BLV users to accurately follow the steps (\textit{e.g.,} describing the use of a measuring spoon, amount of bacon used, and how they're presented). These details are especially important for BLV users, who might otherwise miss visual cues and inaccurately follow the action.

Among baselines, Gemini with the full video context generated the fewest hallucinations but surfaced fewest new information beyond what was already narrated. GPT-4o and LLaVA-OneVision which generated descriptions per step, often included more detailed output -- but much of it focused on irrelevant visual elements like background objects or camera transitions, leading to redundant descriptions between steps.
When given only the step name (without narration), these models tended to extract more visual information not mentioned in the narration, but also hallucinated more due to ambiguity in object recognition (\textit{e.g.,} misidentifying liquids or powders). Providing full narration reduced hallucinations but limited the generation of new facts. Among all models, the open-source LLaVA-OneVision generated the highest rate of hallucinations (Figure~\ref{fig:pipeline_qualitative}).


\begin{table}[]
\resizebox{\columnwidth}{!}{%
\begin{tabular}{@{}llllllll@{}}
\toprule
\multicolumn{2}{l}{}                                                 & \multicolumn{1}{l}{\textbf{Vid2Coach}} & \multicolumn{1}{c}{\textbf{Gemini}} & \multicolumn{2}{l}{\textbf{GPT-4o}} & \multicolumn{2}{l}{\textbf{LLaVA-OV}} \\
\multicolumn{2}{l}{}                                                &                                        & video                               & step        & full        & step         & full         \\ \midrule
\textbf{\# of New Facts}    & \multicolumn{1}{l}{\textbf{$\mu$}}    & \textbf{11.60}                         & 3.80                                & 6.20             & 5.30             & 6.60              & 4.00              \\
(task-relevant)             & \multicolumn{1}{l}{\textbf{$\sigma$}} & 2.91                                   & 2.04                                & 2.04             & 3.16             & 3.69              & 2.00              \\ \midrule
\textbf{\# of Missed Facts} & \multicolumn{1}{l}{\textbf{$\mu$}}    & \textbf{3.80}                          & 8.90                                & 15.80            & 6.50             & 15.90             & 9.60              \\
(task-relevant)             & \multicolumn{1}{l}{\textbf{$\sigma$}} & 2.39                                   & 3.03                                & 9.3              & 4.12             & 6.72              & 3.27              \\ \midrule
\textbf{Hallucinations}     & \%                                     & \textbf{3.92}                          & 7.14                                & 8.82             & 5.71             & 25.58             & 21.13             \\ \bottomrule
\end{tabular}%
}
\caption{We compared the coverage and accuracy of Vid2Coach-generated descriptions to those generated by state-of-the-art VLMs. Vid2Coach descriptions captured more task-relevant facts with fewer hallucinations.}
\vspace{-20pt}
\label{tab:pipeline_results1}
\end{table}

\subsection{Progress Monitoring}
\ipstart{Method}
To monitor task progress in real-time, Vid2Coach uses Gemini with pre-generated criteria derived from how-to videos  (\S\ref{sec:pipeline_monitoring}). For each instructional step, our system classifies incoming video frames as irrelevant, in-progress, or complete, based on how well they align with the expected criteria. To evaluate the effectiveness of this approach, we conduct an ablation study comparing our criteria-based classification against a baseline setup that directly uses the instruction text as context. Beyond Gemini, we also compare the results of CLIP, which is commonly used for measuring visual-textual similarity. To assess generalizability under realistic conditions, we evaluate on 6 action videos across 2 settings: 1) low-quality, narrow field-of-view streams captured using Meta Glasses from our formative study, and 2) high-resolution, wide field-of-view recordings captured with GoPro from prior work~\cite{li2025oscar}. Both sets of videos are recordings of BLV individuals cooking in their own kitchens. As Vid2Coach is designed to support real-time monitoring, we mimic a streaming setup by feeding in frames one at a time rather than feeding the entire segment context to models.

\ipstart{Results}
Table~\ref{tab:pipeline_results2} shows the average per-frame accuracy over 6 actions (90 frames). 
Overall, using a criteria-based approach improved accuracy by grounding classification in context-specific visual expectation rather than generic instruction text.
High-resolution, wide field-of-view videos recorded yield better performance, as they clearly capture the target objects and actions. For punctual actions (\textit{e.g., ``add 1 cup of flour''}), all models struggled to correctly classify the ``complete'' frames. These actions happen quickly and users often instantly shift the camera away from the result -- such as the flour in the bowl -- making it difficult to visually confirm that the step was completed. Durative actions were more reliably detected, with multiple frames showing ongoing progress and completion status. In contrast, iterative actions were often misclassified when the objects being counted (\textit{e.g.,} egg yolks in a bowl) were only partially visible or occluded during the repetitions. CLIP underperformed across all action types due to its limited understanding of complex prompts -- especially those involving negation (\textit{e.g., the butter is in the pan but has not melted into liquid yet.})


\begin{table}[t]
\small\sffamily\def\arraystretch{}\setlength{\tabcolsep}{0.4em}
\centering
\begin{tabular}{llcccc}
\toprule
& & \textbf{Vid2Coach} &  \textbf{Gemini} & \multicolumn{2}{c}{\textbf{CLIP}} \\
Action types & FOV & & instruction  & criteria & instruction\\
\midrule
\multirow{2}{*}{\textbf{Punctual}} & Narrow & \textbf{0.53} & 0.40 & 0.33  & 0.33 \\
& Wide & \textbf{0.73} & 0.47 &  0.33  & 0.33 \\
\midrule
\multirow{2}{*}{\textbf{Iterative}} & Narrow & \textbf{0.60} & 0.20 & 0.33 & 0.33\\
& Wide & \textbf{0.80} & 0.47 & 0.20 & 0.13\\
\midrule
\multirow{2}{*}{\textbf{Durative}} & Narrow & \textbf{1.00} & 0.47 & 0.33 & 0.33 \\
& Wide & \textbf{1.00} & 0.40 & 0.40& 0.40\\
\bottomrule
\end{tabular}
\caption{We compared Vid2Coach's progress monitoring accuracy (Gemini + criteria) to other approaches across action types and FOVs. Criteria-based prompts were more effective for durative actions, while CLIP showed limited gains.}\label{tab:pipeline_results2}
\vspace{-20pt}
\end{table}

\section{User Evaluation}
To evaluate how well ~\sysname{} supports BLV users in following how-to video tasks, we conducted a within-subjects study with 8 BLV participants who completed end-to-end cooking tasks in their own kitchens.  We compared Vid2Coach with a strong baseline of participants' existing practices for following how-to videos. As the baseline, we provided the original how-to video and its transcript, and allowed participants to use any existing human-powered or AI-powered visual assistance applications they typically use.

\subsection{Method}
\ipstart{Participants}
We recruited 8 BLV people (P4-P11 in Table~\ref{tab:blv_participants}) with cooking experience via local chapter of NFB and word of mouth. We recruited people who had prior cooking experience and were able to participate in the study from their own kitchen.
Prior to the study, we asked for participants' vision, cooking experiences, and availability of any accessible kitchen tools to provide as an input to the Vid2Coach's system to generate more relevant tips and workarounds.

\ipstart{Materials}
We selected 3 cooking videos from YouTube (V1, V11-V12 in Table~\ref{apndx:tech_eval_materials}) that provided English narrated instructions by the video author. We used V1 (Chocolate Chip Cookies) for the tutorial on how to use Vid2Coach. The two videos used in the main cooking tasks were V11 (Bread Flapjack) and V12 (Eggs Benedict). We chose less common recipes to ensure participants did not rely on prior knowledge and had to actively follow the system’s instructions and feedback—unlike very familiar dishes like pasta, which many participants already know how to prepare without guidance. Two tasks were similar in terms of length, amount of narration, number of steps, and estimated duration to finish the cooking task. We also chose recipes that meet the dietary requirements of participants and can be completed within 45 minutes.

\ipstart{Baseline}
In the baseline condition, we provided participants with the original how-to cooking video and its transcript, and allowed them to revisit the recipe as many times as they wanted to. We also encouraged them to use any existing AI tools or human visual assistance applications. 4 participants used human assistance applications (\textit{e.g.,} BeMyEyes or AIRA, P4, P6, P8, P10), 7 participants used smart speakers such as (\textit{e.g.,} Amazon Alexa or Google Home, P4, P8-P11) and P8 used Meta AI (smartglasses based assistant) to ask visual questions.

\begin{figure}[t]
  \centering
  \includegraphics[width=\columnwidth]{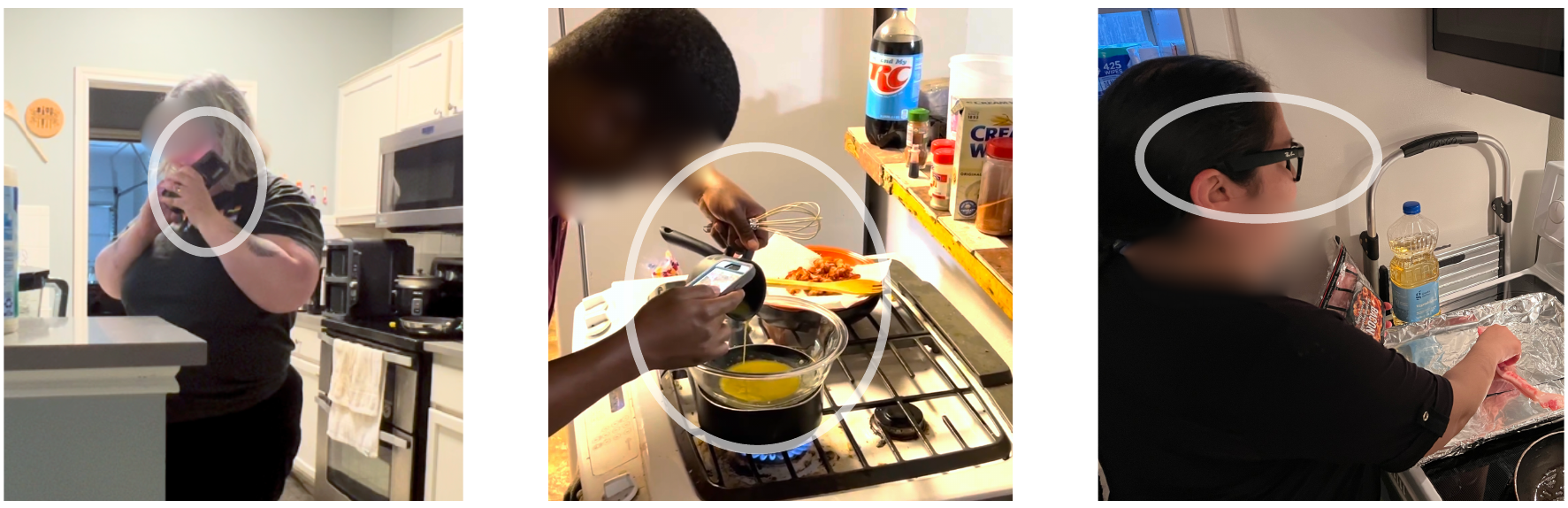}
  \caption{In the baseline condition, participants used their phones to listen to the video instructions (A) or call visual interpreters to get feedback (B). In the Vid2Coach condition, participants wore Meta glasses and used free-form speech to interact with the system. }\label{fig:eval_setup}
\end{figure}

\ipstart{Procedure}
We conducted a 2-hour in-person study\footnote{Approved by the institute's IRB.}. Two researchers visited each participant's home with necessary ingredients and spare kitchen tools. Participants were encouraged to use their own familiar kitchen tools but had the option to use ours if needed. We started the study by collecting demographic and background information about participants' current cooking practices and their use of visual assistance applications. Next, we gave a 10-minute tutorial on how to use \sysname{} with V1. Then, participants completed 2 cooking tasks: one task using the \sysname{} interface and the other task using the baseline condition. In each interface condition, participants were randomly assigned to one of two cooking tasks (V11 or V12). The order of interfaces and recipes was counterbalanced. After each task, we conducted a post-stimulus survey to measure cognitive load with NASA-TLX~\cite{hart1988development} and assess the usefulness of the system features.
At the end of the study, we conducted a semi-structured interview to understand participants' strategies and the strengths and weaknesses of \sysname{}. To analyze task performance, we reviewed recordings, transcripts, and system interaction logs and counted ~\textit{procedural errors} (\textit{e.g.,} missing a step, incorrect order) and ~\textit{technique errors} (\textit{e.g.,} incorrect measurements, incomplete step, incorrect shape, or incorrect use of tools) based on the categorization proposed by Peddi et al.~\cite{peddi2024captaincook4d}. We did not count intentional modifications -- such as deliberate multitasking or strategic workarounds -- and counted only unintentional errors. We got participants' consent for taking and sharing photographs.

\begin{figure*}
  \includegraphics[width=\textwidth]{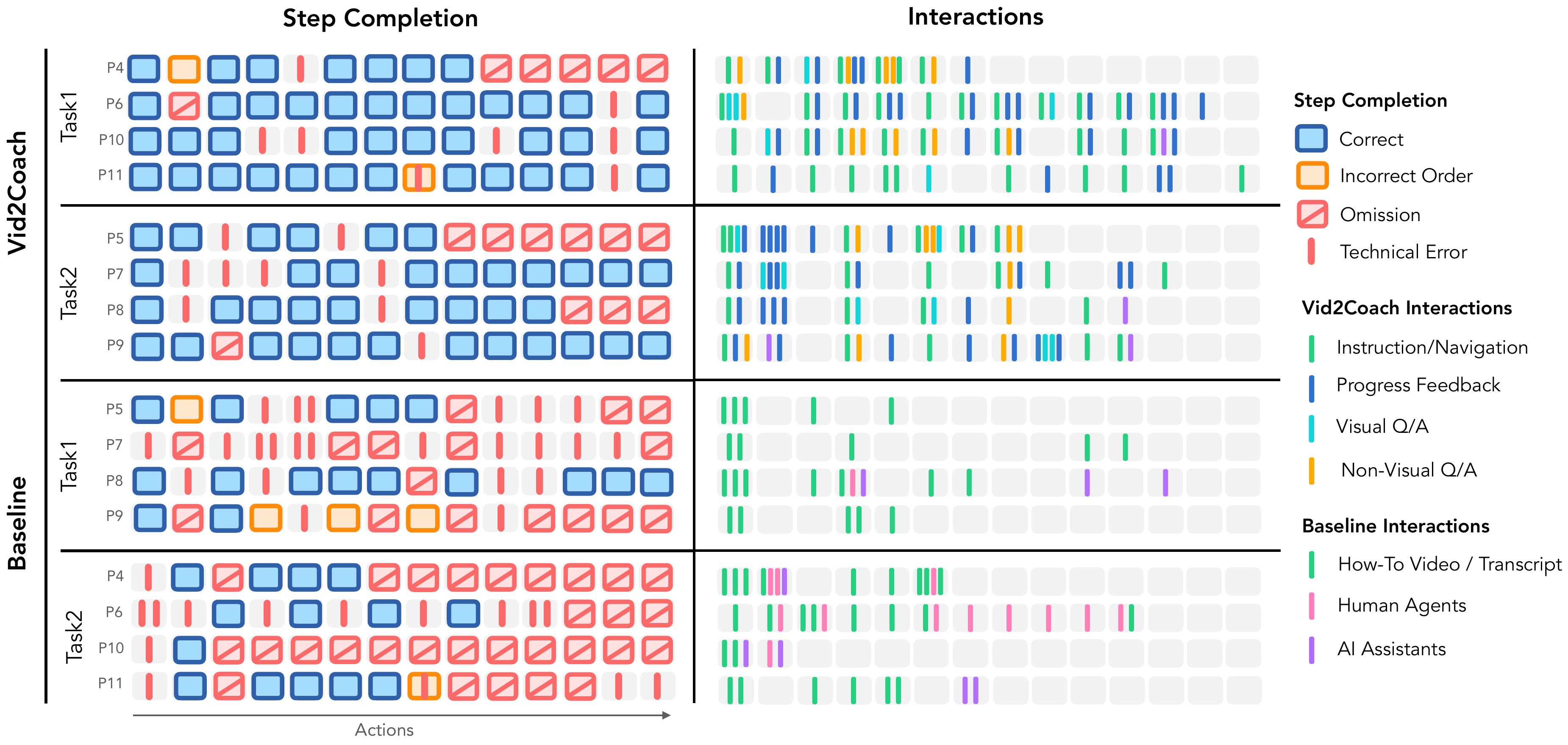}
  \caption{Step completion (left) and user-initiated interactions (right) visualized across participants, grouped by system condition (Vid2Coach vs. Baseline) and task (Task1 vs. Task2). Each row represents a participant, and each column a step in the task. For Vid2Coach, note that we only included user-initiated interactions and not Vid2Coach's consistent feedback for brevity.}
  \label{fig:eval_interactions}
\end{figure*}

\subsection{Task Success and Load}
\begin{figure}
  \includegraphics[width=3.33in]{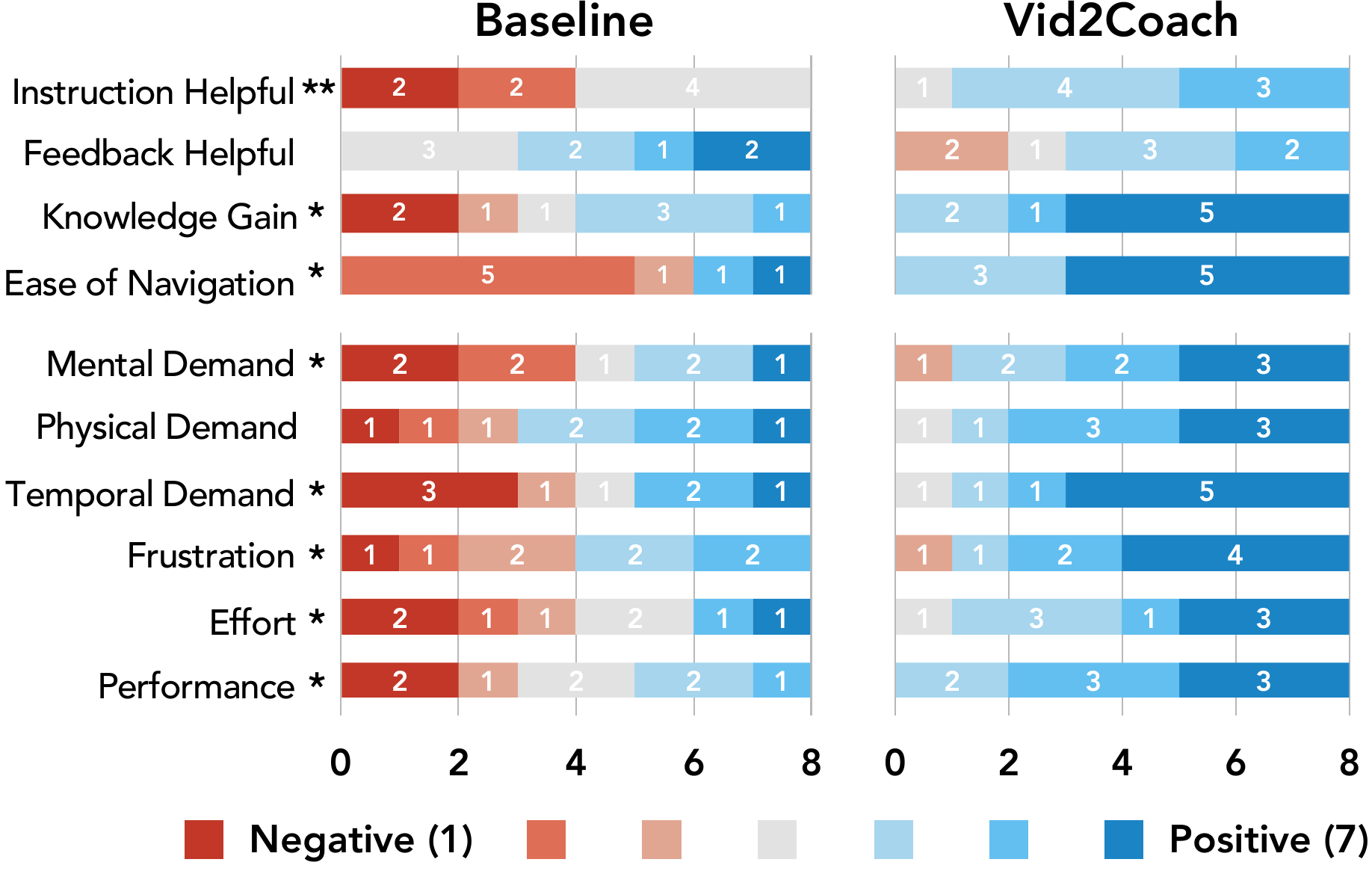}
  \caption{Distribution of the rating scores for the Baseline and Vid2Coach
 (1 = negative, 7 = positive). The asterisks indicate the statistical significance as a result of Wilcoxon text (~\textit{p} < 0.05 is marked with * and ~\textit{p} < 0.01 is marked with **). } 
  \label{fig:eval_survey}
\end{figure}

5 participants using Vid2Coach completed the task, compared to 1 in the baseline condition (Figure~\ref{fig:eval_interactions}).
With Vid2Coach, participants made 58.5\% fewer mistakes ($\mu$=4.38, $\sigma$=2.20 vs. $\mu$=11.00, $\sigma$=3.16; $Z$=2.46; $p$<0.05), showing that the system's real-time guidance contributed to both task completion and accuracy.
As shown in Figure~\ref{fig:eval_survey}, participants using Vid2Coach also experienced significantly lower mental demand ($\mu$=5.75, $\sigma$=1.39 vs. $\mu$=3.38, $\sigma$=2.14; $Z$=2.04; $p$<0.05), temporal demand ($\mu$=6.25, $\sigma$=1.17 vs. $\mu$=3.63, $\sigma$=2.50; $Z$=2.13; $p$<0.05) and frustration ($\mu$=6.00, $\sigma$=1.41 vs. $\mu$=3.88, $\sigma$=1.89; $Z$=1.68; $p$<0.05). While participants in the baseline condition tried to remember a lot of recipe information at once as the navigation is difficult, Vid2Coach users could easily query to repeat or ask for easier workarounds which reduced the mental load.
They also reported higher performance ($\mu$=6.13, $\sigma$=0.83 vs. $\mu$=3.63, $\sigma$=1.85; $Z$=-2.29; $p$<0.05) with lower efforts ($\mu$=5.00, $\sigma$=1.93 vs. $\mu$=3.50, $\sigma$=2.20; $Z$=1.70; $p$<0.05).
Participants explained that the reason for rating high performance is that they have gotten a lot of confirmation from the system on their progress. With the baseline, participants who did not finish all the steps had low confidence and felt more rushed.
While we did not see any significant difference in physical demand ($\mu$=6.00, $\sigma$=1.07 vs. $\mu$=4.38, $\sigma$=2.13; $Z$=1.42; $p$>0.05), 4 participants in the baseline condition mentioned that holding the phone to show the progress to sighted assistants is physically tiring.

    

\subsection{Interaction Behaviors}
We share participants' strategies for utilizing Vid2Coach and how they compare the experience with their current approaches. Figure~\ref{fig:eval_interactions} shows the interaction logs of participants using Vid2Coach and baseline tools. 

\begin{figure}
  \includegraphics[width=3.33in]{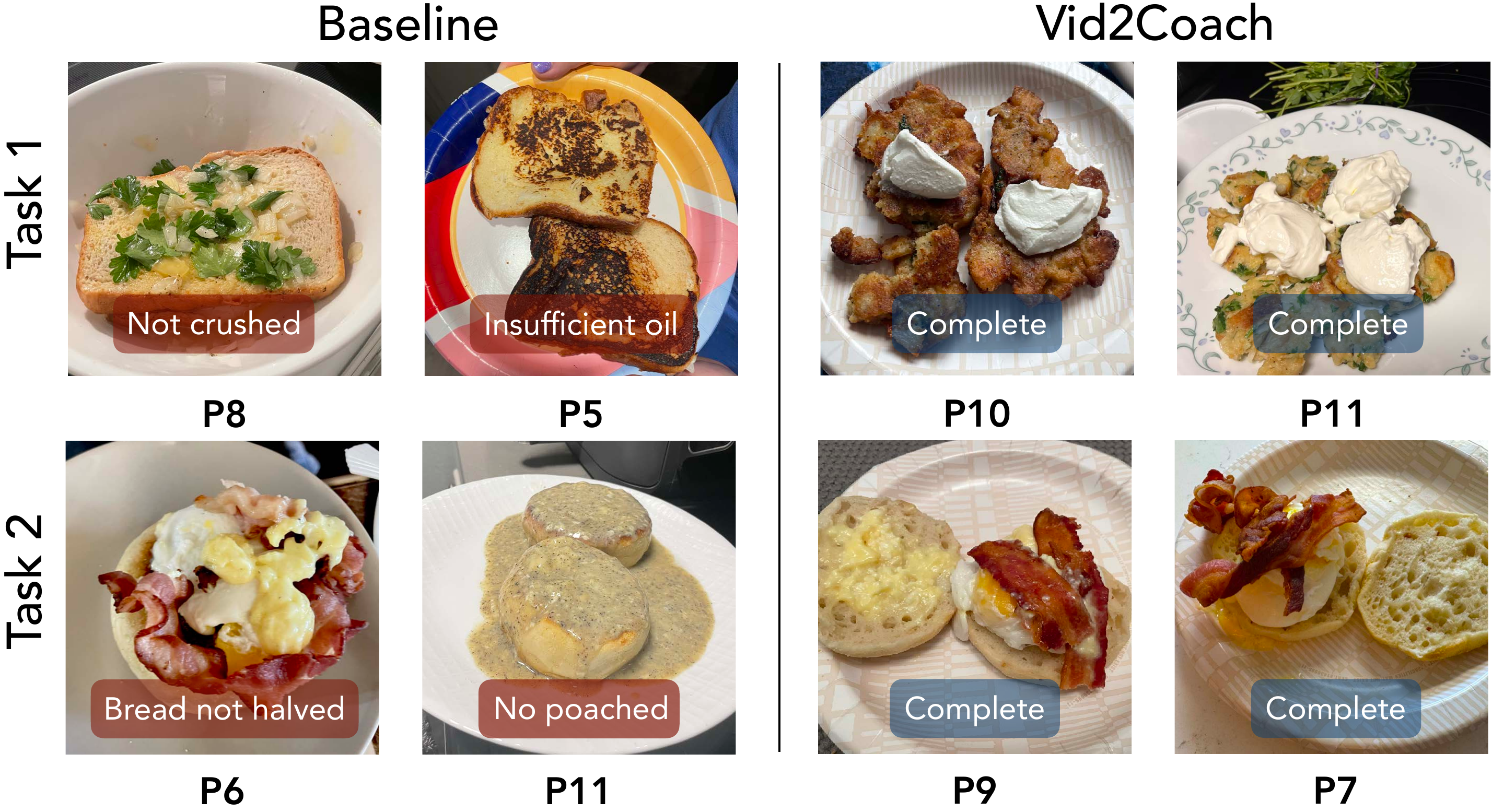}
  \caption{Final dishes from cooking tasks with Vid2Coach and participants' current workflow as baseline. Vid2Coach's detailed instruction and realtime feedback grounded in how-to videos helped participants successfully follow the steps.}
  \label{fig:eval_dishes}
  \vspace{-1.5em}
\end{figure}

\ipstart{Following and Adapting Task Instructions}
Participants rated Vid2Coach’s instructions as significantly more helpful ($\mu$=5.25, $\sigma$=0.71 vs. $\mu$=2.75, $\sigma$=1.39; $Z$=2.46; $p$<0.01) and reported gaining significantly more cooking knowledge ($\mu$=6.50, $\sigma$=0.76 vs. $\mu$=3.75, $\sigma$=1.91; $Z$=2.29; $p$<0.05) compared to the baseline. Although we selected how-to videos with fully narrated instructions, all participants in the baseline condition reported that the original video narration's lack of visual detail reduced their confidence when following instructions. Thus, P5 increased the video volume to infer details about actions based on ~\textit{sound}, such as guessing the tool and chopping technique by listening to the audio. When encountering unfamiliar cooking techniques, participants asked voice assistants (\textit{e.g.,} Alexa and Google Home) or sighted assistance services (\textit{e.g.,} AIRA, BeMyEyes). When P11 asked Alexa for the meaning of the egg poaching, she found the explanations insufficient as Alexa replied ~\textit{``To make a poached egg, you can gently cook it in simmering water by poaching.''} As a result, she mistakenly broke the egg directly into the hollandaise sauce (Figure~\ref{fig:eval_dishes}-P11). P6 called an AIRA agent for guidance on separating egg yolks. Participants frequently made measurement errors or used incorrect tools due to missing visual details in the narration. For instance, both P5 and P8 burned their pancakes due to insufficient oil (Figure~\ref{fig:eval_dishes}-P5), and P4 initially selected a too small pan size and later had to change to a bigger one to fit all bacon pieces.

In contrast, participants using Vid2Coach liked that the instructions is broken down into steps and actions. This granularity reduced the need to repeatedly revisit instructions and allowed them to focus more effectively on cooking. Participants also valued the demonstration details and the optional, personalized tips. P4, who was separating egg yolks for the first time liked the tactile-based explanations -- \textit{``I love that it can tell me easy workarounds. They are not asking me to see it but feel it to check if egg whites all dripped away.''} Similarly, P10 who had limited prior cooking experience, mentioned that he gained practical cooking knowledge applicable to future cooking tasks. P9 highlighted that combining short instructions with demonstration details clarified ambiguitiy -- \textit{``When I am not sure with the instruction, descriptions clarify the direction.''}

\ipstart{Navigating Steps}
Participants rated navigating instructions in Vid2Coach as significantly easier compared to using the baseline video's navigation or transcript ($\mu$=6.25, $\sigma$=1.04 vs. $\mu$=3.25, $\sigma$=2.05; $Z$=2.30; $p$<0.05). In the baseline, most participants first listened to the complete recipe in the baseline condition, either through video (P4-P10) or transcript (P4-P5, P7-P9, P11), and all except P5 revisited the instructions multiple times during the cooking process. Participants generally preferred revisiting instructions through video playback due to easy navigation controls (\textit{e.g.,} rewind or skip by 10 seconds). P5 who did not revisit instructions relied on memory as she preferred not to use her phone while cooking. \textit{``I don't like to touch phones while cooking. I already have to wash hands multiple times whenever I'm touching meat, eggs, so no more.''} However, relying on memory or repeated navigation often led to missed steps even when they are narrated in the video. For example in Figure~\ref{fig:eval_dishes}, both P5 and P8 did not notice that he had to slice the bread, P6 forgot to halve the English muffin, and P5 omitted adding sour cream. Also, P5 chopped the parsley but forgot to put them into the mixture and realized after she finished the cooking task.

When using Vid2Coach, participants enjoyed hands-free voice-based navigation. Although Vid2Coach monitored users' progress for each step and automatically suggested to move on to the next step, participants often paused to control task pacing or manually navigated steps to multitask. For example, P10 manually paused instructions to further finely chop onions as he struggles with sensory sensitivity to vegetables. Also, P5 read the next step instructions to prepare hollandaise sauce ingredients while cooking bacon. P11 attempted to navigate non-linearly using the voice command, ~\textit{``Go back to steps about cutting onions''} but since Vid2Coach only supports navigation to the immediate previous or next step, she had to repeat the "go back" command twice to reach her desired point.


\ipstart{Evaluating the progress with feedback}
Participants appreciated Vid2Coach’s consistent, hands-free feedback while cooking. For example, P4 mentioned that she liked not having to repeatedly ask the same question to check how the bacon looks and not worry about burning it as Vid2Coach keeps giving feedback. Similarly, P9 shared ~\textit{``I like that it keeps describing how I’m getting closer -- it’s like confirmation I’m going in the right direction.''} While some participants preferred consistent guidance initiated by Vid2Coach, a few participants liked that they can also pause the feedback when they are following familiar steps. For instance, P6 said, \textit{``I can count and remember how many bacons I’ve put into the pan, so I didn't need it telling me that there are 2, 3 pieces of bacon.''} Beyond listening to the Vid2Coach's constant feedback, all participants also asked targeted, proactive questions -- such as confirming ingredient size (\textit{e.g., ``Is this the right size of an egg according to the video?''}) or identifying mistakes (\textit{e.g.,} whether the pancake was in the right shape). 
These types of visual questions are typically unanswerable by traditional voice assistants, but Vid2Coach makes them possible by grounding its feedback in the visual content of the how-to video and real-time user progress. As P8 put it, ~\textit{``I never realized, but love that AI can help me assess the doneness visually—I used to think of them [voice assistants] as timers.''}


Despite the value Vid2Coach provided, the difference in participants’ ratings in feedback was not statistically significant ($\mu$=1.28, $\sigma$=4.75 vs. $\mu$=1.28, $\sigma$=5.25; $Z$=0.85; $p$>0.05). 
Still, participants mentioned challenges of relying on human feedback via visual interpreting services. 
While AIRA agents were preferred over BeMyEyes volunteers as they are more trained professionals, they were not always answering the call at the moment and the service was expensive. 3 participants attempted to use AIRA during the study but did not get connected. When P4 asked the agent how to poach an egg, he was unfamiliar with the method and looked up online and called his mom to assist P4, resulting in a delay. Similarly, P6 noted that ~\textit{``Volunteers sometimes don’t know how to cook.''} 
Repeatedly explaining progress to different agents was also frustrating, as P6 added, ~\textit{``I have to tell each agent I’m cooking Benedict and which step I’m on.''} After calling AIRA agents 5 times during the study, P6 eventually asked one to watch the original recipe video and walk him through it step-by-step. Although this was helpful, P6 said, ~\textit{It wouldn’t work in daily life -- after five minutes the call gets expensive, and it’s hard to hold the phone while whisking or slicing.''} 

Participants who did not ask for human feedback in the baseline condition frequently utilized Vid2Coach's feedback during the task. P5, who did not like to use phone while cooking, burned her pancake without any feedback. She noted ~\textit{``It's safe to overcook because it has eggs, and I cannot check when it's brown.''} P7 similarly avoided calling human agents, explaining ~\textit{``I usually don’t use them unless I really, really have to. They’re people—so it’s hard to interrupt or ask the way I want, and even if I’m not satisfied, I have to show thanks.''} Despite not using human assistance, both P5 and P7 actively asked Vid2Coach for feedback throughout their sessions, noting it was relevant, easy to access, and reduced their hesitation. P7 summarized: ~\textit{``I was really impressed with how helpful and quick the feedback was. Only question I have is -- when will this be available?''}

\ipstart{Additional Usecases}
Beyond receiving step-by-step instructions, tips, and feedback, participants used Vid2Coach to ask a variety of contextual questions during cooking. These included object identification (\textit{e.g., ``Which of these is salt?''}), cooking knowledge (\textit{e.g., ``What’s the difference between simmering and boiling?''}), and measurement conversions (\textit{e.g., ``How much is one stick of butter in grams?''}). Some participants also used the system for scene understanding, such as asking ~\textit{``Does it look smoky here?''}, to assess their environment.
Participants familiar with AI visual question answering tools developed specific strategies to increase the accuracy of Vid2Coach’s responses. For example, P5 asked the same question multiple times while scanning different areas around the stove to locate any dropped pieces of bacon. P11 rotated the salt bottle while asking to ensure the label was visible, noting that her hand might otherwise be covering it. 

\subsection{Limitations and Opportunities}
\ipstart{Failure Case Analysis}
We also observed several limitations during deployment, offering insights for future improvments. First, with the limited visual input, a combination of narrow field-of-view and occlusions, Vid2Coach often missed user mistakes or provided incorrect feedback. For instance, Vid2Coach missed bacon pieces that fell to the floor (P5) or repeatedly misjudged pancake doneness by only seeing the top surface while the bottom was already brown (P6). We also observed edge cases where the system’s visual feedback was partially correct but lacked nuance. For example, when some bacon slices were fully cooked while others remained raw, Vid2Coach instructed participants to continue cooking, resulting in burnt pieces. These example cases highlight the need for more fine-grained, localized feedback.

We also observed speech-related failures when participants' commands were unheard due to kitchen noise (\textit{e.g.,} sizzling bacon and running fans), requiring participants to repeat themselves. There were also intent classification errors, such as misinterpreting ~\textit{``It’s fun''} as ~\textit{``It’s done,''} and advancing to the next step.
Participants also pointed out that the system lacked spatial memory and broader situational awareness, not handling questions like ~\textit{``Where did I put the spatula?''} or to assist with actions not directly related to the recipe. For example, P5 unknowingly placed an eggshell on top of a closed trashcan lid, and Vid2Coach did not detect it as mistake because it only tracked for recipe-step related mistakes. Also, even when Vid2Coach provided timely and accurate feedback, some physical tasks—such as flipping pancakes or bacon—remained challenging to perform correctly without visual confirmation.

\ipstart{Future Opportunities}
\camready{Participants often adapted the step sequence by reordering steps for parallel task execution or skipping steps based on dietary preferences. However, this flexibility sometimes led to unintentional skips. In the future, intent can be inferred by analyzing interaction patterns such as rapid, sequential skips (often errors) vs. deliberate reordering, and checking whether the new action sequence still results in a valid completion of the recipe~\cite{idrees2023improved}. Also, Vid2Coach can mark core steps whose omission would compromise safety or outcome (\textit{e.g.,} turning off a burner) and optional steps (\textit{e.g.,} sprinkling cinnamon), and confirm users for skipping core steps. As steps are interdependent, skipping or altering one affect subsequent ones (\textit{e.g.,} leaving bread unsliced changes cooking time). We can leverage step dependencies for flexible guidance.}

Participants also saw Vid2Coach’s potential to enable independent task skill learning in everyday lives. Compared to human assistants, participants found Vid2Coach more readily available, task-relevant, and nonjudgmental. Several participants emphasized the personal motivation and sense of achievement they felt by completing tasks on their own, even when human help was accessible. P4 noted~\textit{``Even if sighted volunteers are available 24/7 for free, I want to know I can achieve it on my own.''} P11, who saw the potential in personalization of Vid2Coach said~\textit{``Unlike human assistants whose description styles change every call, I can train it [Vid2Coach] to be personalized to me.''} P5 also reflected on societal opportunities~\textit{``I think it can help blind students get more jobs. A lot of sighted people assume blind people can’t learn new skills quickly or follow manuals. But with this, we can.''}

Finally, participants envisioned new application areas beyond cooking. These included fashion coaching (\textit{e.g.,} converting outfit videos into accessible styling guidance) and art instruction (\textit{e.g.,} teaching drawing techniques with bind-friendly workarounds). These suggestions emphasize the potential of video-based task assistants to support not only procedural tasks, but also the acquisition of non-procedural knowledge, opening exciting directions for future research in accessible creativity support.

\section{Exploratory Extensions}
While Vid2Coach was designed and evaluated in the context of cooking, we conducted additional 
exploration study in adjacent hands-on tasks, including assembly and decoration, to explore the generalizability of our approach. These domains, like cooking, involve visually-guided, step-based physical manipulation, and present similar accessibility challenges for BLV users. 
We selected two how-to videos for this study: flower arrangement (V13~\cite{flower_video}) and gingerbread house assembly (V14~\cite{gingerbread_video}). P5, who had expressed interest in using Vid2Coach for a wider range of tasks, completed both activities during a 1.5-hour session (30 minutes per task), followed by a semi-structured interview. The materials for both tasks were provided in advance, and P5 had no prior experience with either activity. To tailor the system’s guidance and feedback to the new domains, we adapted Vid2Coach’s pipeline using our arts and crafts accessibility dataset (\S\ref{sec:supplementing_strategies}).

\ipstart{Flower Arrangment}
Compared to cooking, P5 made intentional deviations based on aesthetic preferences rather than frequently replaying instructions. This task's flexibility prompted P5 to ask more contextual questions such as ~\textit{``Where do I put this small flower for better harmony?''} P5 avoided general evaluation questions like ~\textit{``Does this look good?''} in favor of specific queries (\textit{e.g., ``Is there good distinction in height for balance? Does it have many colors?''}) as subjective responses are hard to verify. 
Because the task was less time-sensitive than cooking, she also felt more comfortable using Vid2Coach at a slower pace, explaining that details matter more than speed in this case to inform aesthetic decisions.


\ipstart{Decorative Baking}
When assembling the gingerbread house, P5 faced challenges identifying and positioning pieces even with step-by-step instructions. Unlike cooking's visually distinct ingredients, this task had similarly shaped components that were difficult to differentiate without sight. When P5 asked if a piece of a side wall was for the roof, Vid2Coach incorrectly confirmed, revealing limitations in distinguishing subtle differences. These challenges reflect issues in broader tasks like origami or knitting, where object and action distinctions may be subtle. P5 also highlighted the need for feedback that differentiates between intentional variation and mistakes. For example, repositioning a gingerbread window might be a design choice or an error. Understanding that distinction requires reasoning and user understanding beyond visual correctness.

\vspace{1em}

\begin{figure}[t]
  \centering
  \includegraphics[width=0.8\columnwidth]{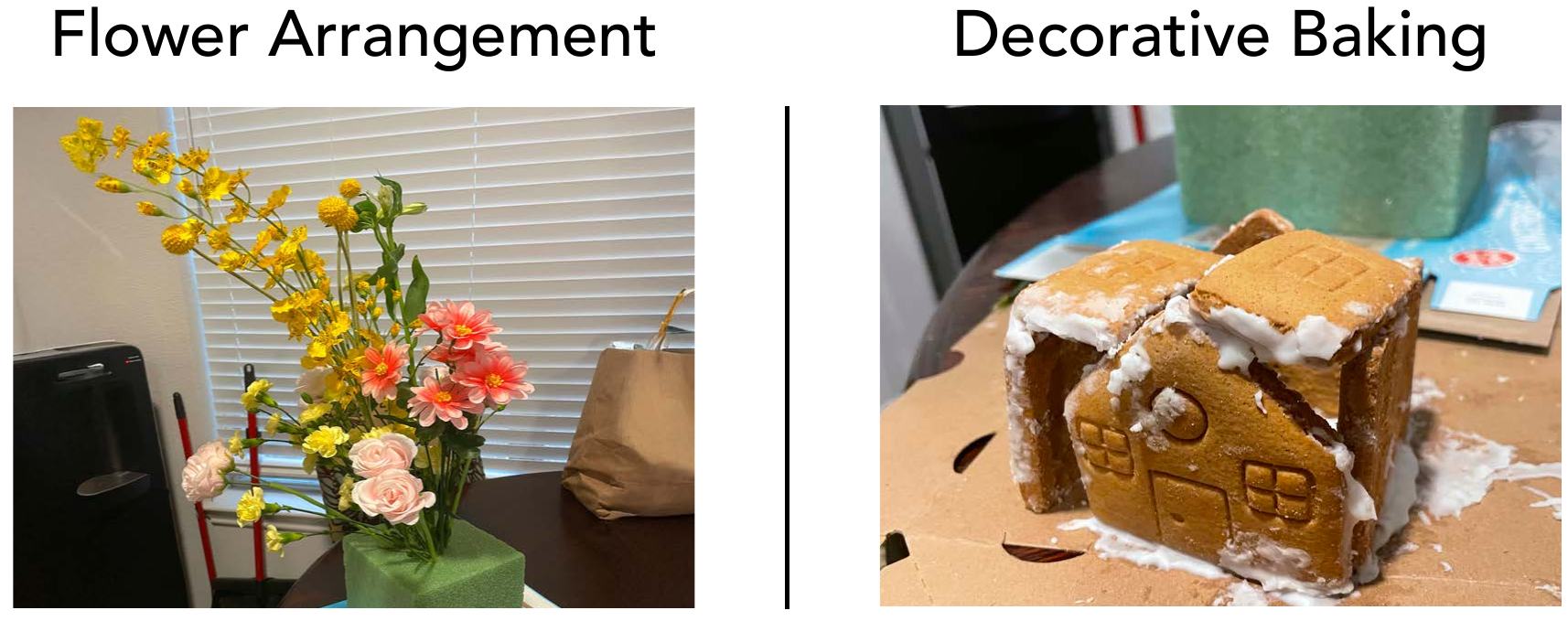}
  \caption{In our extension study, P5 explored the use of Vid2Coach for flower arrangement and decorative baking.}\label{fig:extension}
  \vspace{-1.5em}
\end{figure}

\section{Discussion}









\subsection{Video as a Knowledge Source}

Compared to in-person instruction, how-to videos offer flexible and scalable learning, allowing learners to choose the content and instruction style they prefer and to learn at their own pace. Our work is motivated by how-to videos' potential as learning resources and BLV individuals' interest in accessing them.
Vid2Coach leverages videos in two key ways: transforming them into interactive task assistants and drawing from BLV-shared videos to incorporate accessible practices grounded in lived expertise. This approach allows Vid2Coach to remain anchored in mainstream content while supporting non-visual workflows. While designed for BLV users, Vid2Coach's framework can also benefit sighted users whose hands and eyes are occupied during tasks.

Still, Vid2Coach's guidance effectiveness depends on the original video's quality and completeness. Some videos skip steps or use pre-prepared materials (\textit{e.g.}, pre-cooked vegetables) such that it is not clear how to follow the video.
We did not address narration-less (ASMR-style) videos that rely entirely on visual and audio cues. Making these accessible requires shifting from augmenting narration to interpreting on-screen text and sonic cues as instructions.
Future systems could help BLV users filter videos based on information richness, similar to prior work on accessibility ratings~\cite{liu2021makes}. Vid2Coach could also combine instructions across multiple videos to offer alternatives~\cite{yang2025videomix, ashutosh2024detours}, supporting just-in-time customization (\textit{e.g., ``How can I make this frosting healthier?''}). While Vid2Coach currently adapts mainstream how-to videos to make them more accessible for BLV individuals, future work can encourage creators to consider accessibility from the start~\cite{liu2022crossa11y, peng2021say}.


One reason videos are popular is their ~\textit{immersiveness} -- the chef’s voice, ambient kitchen sounds, and rhythm of actions create a sense of presence. When not using Vid2Coach, participants tried to listen closely to the how-to videos to pick up cues not made explicit in the narration. Future iterations of Vid2Coach could preserve and enhance this immersion by generating synthetic, chef-style feedback using voice cloning, or by integrating audio effects extracted from or synthesized to match the original~\cite{ning2024spica}.


\subsection{Developing Real-World Assistants}
In evaluating Vid2Coach with video streams from Meta smartglasses, we observed key challenges in real-time understanding: models lack full scene context at inference, and video quality is limited by narrow field-of-view and motion blur.
In high-stakes situations like cooking over an open flame, delayed or incorrect feedback can have serious consequences. Similar to WorldScribe~\cite{chang2024worldscribe}, we balanced latency and accuracy using a dual-model approach: one prioritizing low-latency reasoning for immediate feedback, another providing more accurate but slower analysis. Participants confirmed that fast feedback was crucial for time-sensitive cooking tasks, while crafting could tolerate delays for precision.

As smartglasses become more affordable and lightweight, more BLV individuals are incorporating them into daily life. These devices offer hands-free interaction and constant visual access, but they also come with limitations such as short battery life (Meta glasses currently support 50-minutes) and inconsistent framing of target objects. For example, during our study, the system often struggled when the camera view did not capture the object of interest. This highlights the need for spatially-aware feedback like: ~\textit{``Your egg mixture is not visible in the frame—please turn slightly right''}~\cite{liu2024right}. Future systems could combine egocentric camera input from smartglasses with exocentric views (\textit{e.g.,} a kitchen-mounted camera) to provide a more complete sense of the user’s environment.
Smartglasses or AR assistants often leverage intent cues like eye gaze or hand pointing, which are less reliable for BLV people. Future systems can explore alternative intent elicitation, such as tracking hand movements, voice commands, or environmental context.


\camready{\subsection{Expanding to New Activity Domains}}
Participants expressed interest in using Vid2Coach for a variety of tasks beyond cooking, including cleaning, home repair, outfit selection, and makeup. However, tasks that involve fewer objects and less distinctive action names present new challenges. For instance, origami tutorials often use repeated instructions between ~\textit{fold} and ~\textit{crease}, requiring complex manipulation that is hard to convey verbally. Also, dance how-to videos rarely describe movements in detail, relying instead on rhythmic cues like ~\textit{``One, two, three''} and have limited datasets available for segmenting and describing actions. Supporting these broader tasks may require integrating exocentric video perspectives and developing new models to better capture user movement and the environment.

Unlike kitchens where participants were generally comfortable being recorded~\cite{windl2022skewed}, non-cooking scenarios may raise more privacy concerns. Tasks involving the face (\textit{e.g.,} makeup), full body (\textit{e.g.,} dance), or clothing (\textit{e.g.,} fashion) require systems to handle sensitive visual data with care. In these contexts, providing feedback on a person’s appearance goes beyond functional guidance and involves subjective or aesthetic decisions, which can impact users' self-perception and confidence~\cite{li2022feels}.

\camready{\subsection{Scaling Instructional Knowledge}}
While Vid2Coach uses curated accessibility guidelines and BLV-shared practices for common tasks like cooking and crafting, resources for less common tasks (\textit{e.g.,} furniture assembly, planting) are limited. Future systems can incorporate community-sourced knowledge, while also actively eliciting and supporting the sharing of such knowledge. This could involve creating lightweight platforms for BLV users to exchange non-visual workarounds, similar to how blind communities share tips through forums and mailing lists. Peer-to-peer knowledge exchange may surface effective strategies that may not yet be captured in formal accessibility resources.
\camready{To reduce the time and cost for dataset collection, future work can consider LLM-based agents that automatically crawl task-relevant documents and verify credibility~\cite{huang2024can}. Vid2Coach can also refine or expand retrieval queries to better handle steps unseen in the dataset. For example, it can broaden the search by transforming a specific and complicated query into generic sub-queries~\cite{chan2024rq}, or substituting terms with similar ingredients/tools~\cite{lei2024corpus}. For domains like exercise, dance, and the arts, generic written tips are scarce and often less useful than personalized actionable feedback, as the task success depends more on body control and technique. To detect deviations from a reference action and generate concrete correction feedback (\textit{e.g., lift your elbow slightly higher}), future systems can analyze users' live hand- or body-pose data or leverage expert feedback dataset~\cite{ashutosh2024expertaf}. Finally, hybrid models can offer fallback support with human experts (VRTs) and peer-support communities, and learn human responses and automate itself over time~\cite{huang2018evorus}.}

\vspace{10pt}
\subsection{Beyond Voice Assistants}
While Vid2Coach currently provides verbal guidance, future systems could expand beyond voice to support a wider range of sensory and embodied interactions for task assistance. For instance, augmented reality (AR) overlays could benefit low vision users. Similar to CookAR~\cite{lee2024cookar}, which used visual cues to support safe interaction with kitchen tools, Vid2Coach could use visual overlays to show task progress (\textit{e.g.,} using a saturated green overlay to indicate completion or red to flag mistakes) or highlight relevant ingredients during each step.
Multimodal feedback, such as sound cues or haptic feedback using a smartwatch, could provide timely, attention-grabbing feedback for urgent events (\textit{e.g.,} detecting proximity to a hot pan). Our study participants found it difficult to follow certain actions purely based on verbal descriptions (\textit{e.g.,} flipping a pancake or knowing the right whisking speed).
Tactile or directional force feedback~\cite{narayanan2022enabling} could convey these nuanced physical motions more effectively.
Finally, embodied agents like robots can help with less-creative parts of a task, such as chopping or cleaning. Such support will require new exploration into how to support control, trust, and verification of agents' actions for BLV people.


\subsection{Complementary Expert-AI Support}
\camready{The field of Virtual Reality Therapy (VRT) is relatively small and growing more slowly than the increasing demand~\cite{connors2023perspectives}. As VRTs teach a wide range of life skills (\textit{e.g.,} navigation, technology), most of their time goes to foundational, in-person training, leaving little capacity for personalized recipe coaching (Section~\ref{study_findings}). Vid2Coach addresses this gap by supporting BLV users in independently learning new recipes after they have received foundational training (\textit{e.g.,} holding a knife) from VRTs. VRTs in our formative study were excited for such systems, recognizing their potential to extend support between in-person sessions. Future HCI work can explore how VRTs and systems like Vid2Coach can collaborate more effectively, supporting long-term training programs. 
Just as in physical therapy or music lessons --- where basics are taught in person and home exercises are assigned~\cite{argent2018patient} --- future VRTs can give basic training, assign recipe videos matched to students’ skill level, and review the user’s cooking session to offer nuanced coaching and encouragement. Systems can flag challenge points (\textit{e.g.,} moments of hesitation or unsafe motion)~\cite{mysore2018porta} and let the VRT select or record brief corrective clips that the system bundles into a learner-specific practice set. In this workflow, the system handles detection and summarization, while the VRT provides the expert guidance, blending automated detection with human expertise in complementary roles.}

\section{Conclusion}
Vid2Coach bridges the accessibility gap in how-to videos by transforming them into wearable, context-aware assistants that provide accessible instructions and real-time feedback to BLV users. Grounded in our observational studies with VRTs providing remote guidance to BLV individuals, we designed Vid2Coach to offer both proactive and responsive support, grounded in how-to videos and accessibility resources. Our study demonstrates that BLV users can complete tasks more accurately and confidently with Vid2Coach, and all expressed a desire to use it in daily life. Our work demonstrates how AI systems can enable more sclable, flexible, and independent skill learning for BLV individuals in real-world setting.
\begin{acks}
Mina Huh is supported by a Google Ph.D. fellowship.
\end{acks}

\bibliographystyle{ACM-Reference-Format}
\bibliography{sample-base}
\clearpage
\onecolumn
\appendix


\section{PARTICIPANTS}

\begin{table*}[htbp!]
\small\sffamily\def\arraystretch{1}\setlength{\tabcolsep}{0.8em}
    \centering
    \begin{tabular}{llllll}
        \toprule
       PID  & Gender & Age & Job & Teaching Experience & Trainings Provided \\
       \midrule
        VRT1 & Female & 53 & Certified Occupational Therapist & 18 years & Cooking, Cane travel, Public Transportation \\ 
        VRT2 & Female & 28 & Vision Rehab Therapist & 6 years & Cooking, Technology, Communication, Social etiquette \\ 
        VRT3 & Female & 28 & Teacher of BLV & 4 years & Cooking, House cleaning, Shopping \\ 
        \bottomrule
    \end{tabular}
    \caption{Demographics of VRT participants in the formative study} 
    \label{tab:vrt_participants}
\end{table*}

\begin{table*}[htbp!]
\small\sffamily\def\arraystretch{1}\setlength{\tabcolsep}{0.8em}
    \centering
    \begin{tabular}{lllllll}
        \toprule
       PID  & Gender & Age & Job & Visual Impairment & Current Vision Duration & Cooking Experience  \\
       \midrule
        P1 & Female & 29 & Teacher of BLV & Totally Blind & 27 years &  18 years \\ 
        P2 & Female & 51 & Accessibility Tester & Totally Blind & 14 years &  30 years \\ 
        P3 & Male & 40 & Vocational Rehab Teacher & Totally Blind & 5 years &  30 years \\ 
        P4 & Female & 37 & Massage Therapist & Totally Blind & 5 years &  1-2 years \\ 
        P5 & Female & 36 & Teacher & Low Vision & 30 years &  20 years \\ 
        P6 & Male & 36 & Teacher & Totally Blind & 20 years &  18 years \\ 
        P7 & Female & 30 & Substitute Teacher & Totally Blind & 9 years &  6 years \\ 
        P8 & Female & 59 & Teacher & Totally Blind & 40 years &  30 years \\ 
        P9 & Female & 37 & Unemployed & Totally Blind & 37 years &  20 years \\ 
        P10 & Male & 21 & Unemployed & Some Light/Shadow Perception & 21 years &  3 years \\ 
        P11 & Female & 36 & Opera Singer & Totally Blind & >10 years & 18 years \\ 
        \bottomrule
    \end{tabular}
    \caption{Demographics of BLV participants (Formative study: P1-P3, User Evaluation: P4-P11, Extension Study: PX)} 
    \label{tab:blv_participants}
\end{table*}

\section{Study Materials}\label{apndx:tech_eval_materials}

\begin{table}[h!]
\small\sffamily\def\arraystretch{0.95}\setlength{\tabcolsep}{0.4em}
  \centering
  \begin{tabular}{cccc}
    \hline
    \textbf{Video ID} & \textbf{Duration} & \textbf{Task} & \textbf{URL} \\ \hline
    V1 & 6:00 & Chocolate Chip Cookies & \cite{v1} \\ \hline
    V2 & 4:35 & Eggs Benedict & \cite{v2} \\ \hline
    V3 & 4:01 & Mini Pavlovas & \cite{v3} \\ \hline
    V4 & 5:06 & Tortilla Pizza & \cite{v4} \\ \hline
    V5 & 8:30 & Strawberry Jam & \cite{v5} \\ \hline
    V6 & 5:54 & Dumpling & \cite{v6} \\ \hline
    V7 & 6:56 & Omelette & \cite{v7} \\ \hline
    V8 & 3:59 & Beef and Broccoli & \cite{v8} \\ \hline
    V9 & 5:06 & Mashed Potatoes & \cite{v9} \\ \hline
    V10 & 7:28 & Tiramisu & \cite{v10} \\ \hline
    V11 & 3:54 & Bread Flapjack & \cite{v12} \\ \hline
    V12 & 3:19 & Eggs Benedict & \cite{v11} \\ \hline
    V13 & 9:04 & Flower Assembly & \cite{flower_video} \\ \hline
    V14 & 4:06 & Gingerbread House & \cite{gingerbread_video} \\ \hline
  \end{tabular}
  \caption{Video Materials (V1-V3: Observational Study, V1-V10: Pipeline Evaluation, V11-V12: User Evaluation, V13-V14: Extension }
  \label{tab:video_materials}
\end{table}

\subsection{Experts Observational Study}
In Section~\ref{sec:formative_method}, we conducted a conversational analysis~\cite{goodwin1990conversation} with the video recordings and annotated the transcripts. Each utterance was attributed to either the BLV or VRT participant and segmented at the sentence level. Sentences that formed a single cohesive instruction (\textit{e.g., ``You are going to add flour. One and a half cups of flour.''}) were merged into a single utterance. Unrelated speech, such as casual chatter, was excluded from annotation. Two researchers collaboratively developed a coding scheme by reviewing a representative subsample from each session and refining the labels and definitions through discussion. One researcher then annotated the full dataset, and a second researcher independently reviewed the annotations to discuss and resolve any conflicts (See Supplementary Material). 
To structure our analysis, we applied 4 labels -- \textit{instruction}, \textit{supplementary} (tips or workarounds), and \textit{progress description} (on BLV participants' current task state) and \textit{question} -- which helped us identify patterns of support and information exchange. We did not use a separate label for responses; instead, they were categorized as instructions, supplementary information, progress descriptions, or questions based on their content.

\section{Pipeline Prompts}\label{sec:pipeline_prompts}

\begin{table*}[!h]
\scriptsize
\resizebox{0.9\textwidth}{!}{%
\begin{tabular}{@{}p{15cm}@{}}
\toprule
\texttt{Given the definitions of the taxonomy, classify the provided sentence into one of the eight categories:} 
\texttt{[Greeting, Overview, Method, Supplementary, Explanation, Description, Conclusion, and Miscellaneous].} 
\texttt{Do not add sub category.}

\texttt{1. Greeting} \\
\texttt{Opening: Starting remarks and instructor/channel introductions.} \\
\texttt{Example: "Hey, what's up you guys, Chef [...] here."} \\
\texttt{Closing: Parting remarks and wrap-up.} \\
\texttt{Example: "Stay tuned, we'll catch you all later."} \\\\

\texttt{2. Overview} \\
\texttt{Goal: Main purpose of the video and its descriptions.} \\
\texttt{Example: "Today, I'll show you a special technique which is totally special and about image pressing."} \\
\texttt{Motivation: Reasons or background information on why the video was created.} \\
\texttt{Example: "[...] Someone is making a very special valentine's day meal for another certain special someone."} \\
\texttt{Briefing: Rundown of how the goal will be achieved.} \\
\texttt{Example: "I'm pretty sure that just taking a pencil and putting it over the front and then putting a bunch of rubber bands around the pencil [...] that's going to do it."} \\\\

\texttt{3. Method} \\
\texttt{Subgoal: Objective of a subsection.} \\
\texttt{Example: "Now for the intricate layer that will give me the final webbing look."} \\
\texttt{Instruction: Actions that the instructor performs to complete the task.} \\
\texttt{Example: "We're going to pour that into our silicone baking cups."} \\
\texttt{Tool: Introduction of the materials, ingredients, and equipment to be used.} \\
\texttt{Example: "I'm also going to use a pair of scissors, a glue stick, some fancy tape or some regular tape."} \\\\

\texttt{4. Supplementary} \\
\texttt{Tip: Additional instructions or information that makes instructions easier, faster, or more efficient.} \\
\texttt{Example: "I find that it's easier to do just a couple of layers at a time instead of all four layers at a time."} \\
\texttt{Warning: Actions that should be avoided.} \\
\texttt{Example: "I don't know but I would say avoid using bleach if you can."} \\\\

\texttt{5. Explanation} \\
\texttt{Justification: Reasons why the instruction was performed.} \\
\texttt{Example: "Because every time we wear our contact lenses, makeup and even dirt particles [...] might harm our eyes directly."} \\
\texttt{Effect: Consequences of the instruction.} \\
\texttt{Example: "And these will overhang a little to help hide the gap."} \\\\

\texttt{6. Description} \\
\texttt{Status: Descriptions of the current state of the target object.} \\
\texttt{Example: "Something sticky and dirty all through the back seat."} \\
\texttt{Context: Descriptions of the method or the setting.} \\
\texttt{Example: "[...] The process of putting on a tip by hand [...] takes a lot of patience but it can be done if you're in a pinch."} \\
\texttt{Tool Specification: Descriptions of the tools and equipment.} \\
\texttt{Example: "These are awesome beans, creamy texture, slightly nutty loaded with flavor."} \\\\

\texttt{7. Conclusion} \\
\texttt{Outcome: Descriptions of the final results of the procedure.} \\
\texttt{Example: "And now we have a dinosaur taggy blanket that wrinkles, so a fun gift for any baby on your gift giving list."} \\
\texttt{Reflection: Summary, evaluation, and suggestions for the future about the overall procedure.} \\
\texttt{Example: "However, I am still concerned about how safe rubbing alcohol actually is to use so maybe next time, I will give vodka a try."} \\\\

\texttt{8. Miscellaneous} \\
\texttt{Side Note: Personal stories, jokes, user engagement, and advertisements.} \\
\texttt{Example: "Tristan is back from basketball - He made it on the team so it's pretty exciting."} \\
\texttt{Self-promotion: Promotion of the instructor of the channel (i.e. likes, subscription, notification, or donations).} \\
\texttt{Example: "So if you like this video, please give it a thumbs up and remember to subscribe."} \\
\texttt{Bridge: Meaningless phrases or expressions that connect different sections.} \\
\texttt{Example: "And we're going to go ahead and get started."} \\
\texttt{Filler: Conventional filler words.} \\
\texttt{Example: "Whoops."} \\\\

\texttt{EXAMPLES:}\\
\texttt{Sentence: Hey, I'm John Kanell.} \\
\texttt{Category: Greeting} \\\\

\texttt{Sentence: And today on Preppy Kitchen, we're making some quick and delicious cranberry orange muffins.} \\
\texttt{Category: Overview} \\\\
\bottomrule
\end{tabular}
}
\caption{Example prompt for information type classification.}
\label{tab:info_type_prompt}
\end{table*}

\begin{table*}[!h]
\scriptsize
\resizebox{0.9\textwidth}{!}{%
\begin{tabular}{@{}p{15cm}@{}}
\toprule
\texttt{This is a transcript of a tutorial video: "\{transcript\_data\}".} \\
\texttt{This is the metadata for this tutorial: "\{metadata\}".} \\
\texttt{Prioritize metadata if the images look different than the metadata.} \\

\texttt{Output a JSON file that segments this into high-level steps. For each step, include:} \\
\texttt{- \textbf{step\_name}} \\
\texttt{- \textbf{actions}: a list of action objects containing:} \\
\quad \texttt{- instruction (single atomic verb)} \\
\quad \texttt{- supplementary (relevant tips, warnings, explanations)} \\
\quad \texttt{- start and end times} \\

\texttt{Use entries with the \textbf{method} role as the main instruction. Supplement with other roles (tip, warning, explanation, etc).} \\
\texttt{Make instructions specific and actionable: include measurements (e.g., "Add 1.5 cups of ...") and tools ("Mix using a spatula ...").} \\

\texttt{\textbf{Important:}} \\
\texttt{- Each instruction must be a clear, single-sentence action centered around one verb.} \\
\texttt{- Split instructions with multiple actions (e.g., “Add sugar and whisk” → two separate actions).} \\
\texttt{- Split iterative actions over different materials (e.g., “Add salt, sugar, and vanilla extract” → three actions).} \\
\texttt{- Merge only if instructions describe the same event.} \\
\texttt{- If multiple actions are in one sentence, assign the same timestamp to each.} \\
\texttt{- Some steps may have no actions if no method-role content is present.} \\
\texttt{- A step’s start time = first action’s start; end time = last action’s end; next step starts at previous step’s end.} \\

\texttt{Also include:} \\
\texttt{- \textbf{tools}: all tools used in this step} \\
\texttt{- \textbf{materials}: all materials/ingredients used in this step} \\
\texttt{- \textbf{new\_tools}, \textbf{new\_materials}: any tools/materials not used in the previous step} \\

\texttt{Do not hallucinate. Only use provided information.} \\

\texttt{Example step:} \\
\texttt{Step(} \\
\quad \texttt{step\_name="Prepare Cookie Dough",} \\
\quad \texttt{actions=[} \\
\quad\quad \texttt{Action(} \\
\quad\quad\quad \texttt{instruction="Add 1 cup of flour into the bowl.",} \\
\quad\quad\quad \texttt{supplementary=["Use precise measurements for the best results."],} \\
\quad\quad\quad \texttt{start=0.0,} \\
\quad\quad\quad \texttt{end=5.0} \\
\quad\quad \texttt{),} \\
\quad\quad \texttt{Action(} \\
\quad\quad\quad \texttt{instruction="Mix the mixture with a spatula until no residue flour is visible.",} \\
\quad\quad\quad \texttt{supplementary=["Hold the bowl with the other hand for stability."],} \\
\quad\quad\quad \texttt{start=5.0,} \\
\quad\quad\quad \texttt{end=10.0} \\
\quad\quad \texttt{),} \\
\quad\quad \texttt{Action(} \\
\quad\quad\quad \texttt{instruction="Let the dough rest for 30 minutes.",} \\
\quad\quad\quad \texttt{supplementary=["Resting the dough helps improve the texture of the cookies."],} \\
\quad\quad\quad \texttt{start=10.0,} \\
\quad\quad\quad \texttt{end=40.0} \\
\quad\quad \texttt{)} \\
\quad \texttt{],} \\
\quad \texttt{tools=["Cup", "Spatula", "Mixing bowl"],} \\
\quad \texttt{materials=["Flour"],} \\
\quad \texttt{new\_tools=["Spatula"],} \\
\quad \texttt{new\_materials=["Flour"],} \\
\quad \texttt{start=0.0,} \\
\quad \texttt{end=40.0} \\
\texttt{)} \\
\bottomrule
\end{tabular}
}
\caption{Pipeline prompt for identifying steps and actions in how-to videos.}
\label{tab:step_segmentation_prompt}
\end{table*}

\begin{table*}[!h]
\scriptsize
\resizebox{0.9\textwidth}{!}{%
\begin{tabular}{@{}p{15cm}@{}}
\toprule
\texttt{Generate response to the following query with the given context. If there is no relevant information, say ``I don't know''.} \\
\texttt{User\_info: \{user\_info\}} \\
\texttt{Query: User is currently performing \{action\}, what are useful tips and workarounds?} \\
\texttt{Context: \{context\} \# Relevant text chunk found across Top-3 documents based on user query} \\
\texttt{Response:} \\
\bottomrule
\end{tabular}
}
\caption{Retrieval augmented generation prompt used for supplementing accessible strategies.}
\label{tab:tip_workaround_prompt}
\end{table*}

\begin{table*}[!h]
\scriptsize
\resizebox{0.9\textwidth}{!}{%
\begin{tabular}{@{}p{15cm}@{}}
\toprule
\texttt{This is information about a tutorial video: "\{step\}" . Output a JSON that consists of the following attributes:} \\
\texttt{tools, materials, actions.} \\

\texttt{For each action, specify an action type between: punctual, iterative, and durative.} \\
\texttt{Punctual actions are brief and occur at a specific moment (e.g., "Put 1 cup of flour").} \\
\texttt{Iterative actions involve repetition or multiple quantities (e.g., "Add 2 rounded teaspoons").} \\
\texttt{Durative actions extend over time and involve continuous motion (e.g., "Whisk the mixture").} \\

\texttt{For each action, specify:} \\
\texttt{in\_progress\_criteria — visual indicators the action is ongoing;} \\
\texttt{completion\_criteria — visual signs that the action is finished;} \\
\texttt{mistake\_criteria — possible visual errors;} \\
\texttt{nonvisual\_completion\_criteria — (optional) sensory cues for completion (e.g., "feels crispy").} \\

\texttt{Note:} \\
\texttt{- Punctual actions should not include in\_progress\_criteria.} \\
\texttt{- completion\_criteria should be grounded in the instruction (e.g., "until brown").} \\
\texttt{- in\_progress\_criteria should not overlap with completion\_criteria.} \\
\texttt{- Only use information provided. Do not hallucinate.} \\

\texttt{Also, extract tools and materials used in this step. If available, include precise amounts from:} \\
\texttt{1/2 cup white sugar, 1/2 cup dark brown sugar, 1 egg, 1 tsp vanilla, 1/2 tsp salt (kosher),} \\
\texttt{1/2 tsp baking soda, 1 1/3 cups AP flour, 1 cup large chocolate chips.} \\

\texttt{\# Example instantiation:} \\

\texttt{example\_step = Step(} \\
\texttt{~~~~tools=['whisk', 'bowl'],} \\
\texttt{~~~~materials=['1 1/3 cups AP flour', '1/2 cup white sugar', 'butter'],} \\
\texttt{~~~~actions=[} \\
\texttt{~~~~~~~~Action(} \\
\texttt{~~~~~~~~~~~~instruction='Put 1 cup of flour into the bowl.',} \\
\texttt{~~~~~~~~~~~~video\_description='The person scoops all-purpose flour into a shiny stainless steel 1-cup measuring cup...'} \\
\texttt{~~~~~~~~~~~~type='punctual',} \\
\texttt{~~~~~~~~~~~~completion\_criteria=['The flour is visible in the bowl.'],} \\
\texttt{~~~~~~~~~~~~mistake\_criteria=['Flour spills outside the bowl.']}, \\
\texttt{~~~~~~~~),} \\
\texttt{~~~~~~~~Action(} \\
\texttt{~~~~~~~~~~~~instruction='Add 3 eggs into the mixture.',} \\
\texttt{~~~~~~~~~~~~video\_description='The person gently cracks three fresh eggs...'} \\
\texttt{~~~~~~~~~~~~type='iterative',} \\
\texttt{~~~~~~~~~~~~in\_progress\_criteria=['One or two eggs are visible in the bowl, but not all three.'],} \\
\texttt{~~~~~~~~~~~~completion\_criteria=['All three eggs are visible in the bowl.'],} \\
\texttt{~~~~~~~~~~~~mistake\_criteria=['More than three eggs added', 'Eggshell is visible.']}, \\
\texttt{~~~~~~~~),} \\
\texttt{~~~~~~~~Action(} \\
\texttt{~~~~~~~~~~~~instruction='Whisk the mixture until it is smooth.',} \\
\texttt{~~~~~~~~~~~~video\_description='The person holds a ceramic bowl steady with one hand while whisking...'} \\
\texttt{~~~~~~~~~~~~type='durative',} \\
\texttt{~~~~~~~~~~~~in\_progress\_criteria=['The whisk is moving through the mixture.'],} \\
\texttt{~~~~~~~~~~~~completion\_criteria=['The mixture looks smooth and consistent.'],} \\
\texttt{~~~~~~~~~~~~nonvisual\_completion\_criteria=['Mixture feels smooth to the touch.'],} \\
\texttt{~~~~~~~~~~~~mistake\_criteria=['Mixture is lumpy or too runny.']}, \\
\texttt{~~~~~~~~)} \\
\texttt{~~~~]} \\
\texttt{)} \\

\bottomrule
\end{tabular}
}
\caption{Pipeline prompt for classifying actions into punctual, iterative, durative actions and generating completion criteria.}
\label{tab:structured_annotation_prompt}
\end{table*}

\clearpage
\section{User Study Example}
\begin{figure*}[htbp!]
  \includegraphics[width=0.95\textwidth]{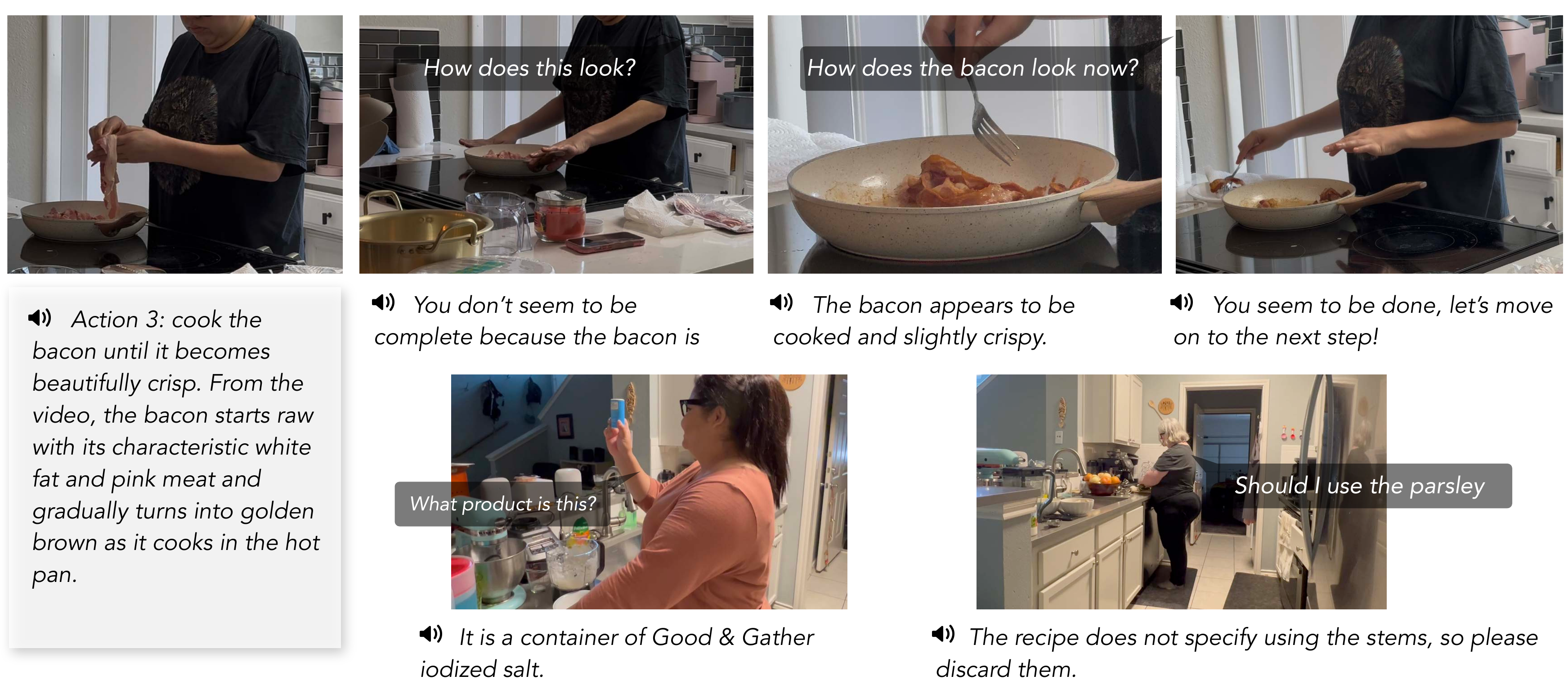}
  \caption{In the user study, participants used Vid2Coach to receive real-time feedback and ask free-form questions during cooking, helping them assess food readiness and ingredient use with confidence.}
  \label{fig:apndx_teaser}
\end{figure*}

\end{document}